\documentclass[epj,final]{svjour}

\usepackage[dvips]{graphicx}
\usepackage{pstricks}
\usepackage[english]{babel}
\usepackage{amssymb,latexsym,amsmath}
\usepackage{cite}
\usepackage{dsfont}
\usepackage{psfrag}
\usepackage{bm} 
\usepackage{multirow}
\usepackage{booktabs}

\def\Id{{\rm 1\kern-.3em I}}
\newcommand{\dslash}{\!\!\!/}

\begin{document}

\definecolor{darkblue}{rgb}{0,0,.5}
\definecolor{darkgreen}{rgb}{0,.5,0}
\definecolor{darkred}{rgb}{.5,0,0}

\title{
  Effects of a spin-flavour dependent interaction on the baryon mass spectrum
}
\author{
  Michael Ronniger\thanks{e-mail: \texttt{ronniger@hiskp.uni-bonn.de}} 
  \and 
  Bernard Ch. Metsch
}

\titlerunning{
  Effects of a spin-flavour dependent interaction 
  on the baryon mass spectrum
}
\authorrunning{
  M. Ronniger \emph{et al.}
}
\institute{
  Helmholtz Institut f\"ur Strahlen-- und Kernphysik (Theorie), 
  Universit\"at Bonn, 
  Nu\ss allee 14-16, 
  D-53115 Bonn,
  Germany
}
\date{\today}

%
\abstract{
The effective quark interaction in a relativistically covariant constituent
quark model based on the Salpeter equation is supplemented by an extra
phenomenological flavour dependent force in order to account for some
discrepancies mainly in the description of excited negative parity $\Delta$
resonances. Simultaneously an improved description of some other features of
the light-flavoured baryon mass spectrum and of some electromagnetic form
factors is obtained.
} 

\maketitle

\renewcommand{\figurename}{Fig.} 

\section{Introduction\label{intro}}
\label{sec:Intro}

The description of the hadronic excitation spectrum re\-mains a major
challenge in strong interaction theory. In spite of recent progress in
unquenched lattice QCD access to excited states is still very
limited~\cite{Edwards,Huey-Wen}. Therefore it seems worthwhile to improve
upon constituent quark model descriptions, which in view of the light quark\linebreak
masses (even taken as effective constituent masses) have to be formulated in
terms of relativistically covariant equations of motion. About a decade ago we
formulated such a quark model for baryons, see
\cite{LoeMePe1,LoeMePe2,LoeMePe3} on the basis of an instantaneous formulation
of the Bethe-Salpeter equation. In this model the quark interactions reflect a
string-like description of quark confinement through a confinement potential
rising linear\-ly with interquark distances as well as a spin-flavour
dependent interaction on the basis of instanton effects, which explains the
major spin-dependent splittings in the baryon spec\-trum.
 
Such a model description should offer an efficient description of masses
(resonance positions), static properties such as magnetic moments, charge
radii, electroweak amplitudes (form factors and helicity amplitudes) with only
a few model parameters. As such they also offer a framework which can be used
to judge in how far certain features could be considered to be exotic.  This
concerns \emph{e.g.} phenomenological evidence for states with properties that
can not be accounted for in terms of excitations of quark degrees of freedom,
as is at the heart of any constituent quark model, but instead requires
additional degrees of freedom as \emph{e.g.} reflected by hadronic
interactions.

A satisfactory description of the major features in the light-flavoured
baryonic mass spectrum could indeed be obtained. These include
\begin{itemize}
\item 
  the linear Regge trajectories with an universal slope for all flavours
  including states up to total angular momenta of $J=\frac{15}{2}$ and excitation
  energies up to 3 GeV, see \cite{LoeMePe2,LoeMePe3};
\item
  the position of the Roper-resonance and three other positive parity excited
  nucleon states well below all other states of this kind. These can be largely
  accounted for by the instan\-ton-induced force, the\linebreak strength of which was
  chosen to reproduce the ground state $N-\Delta$ splitting
  \cite{LoeMePe2,LoeMePe3};
\item
  a plethora of electroweak properties which can be explained without
  introducing any additional parameters, see \cite{Merten,vCauteren,Haupt}.
\end{itemize}

Nevertheless some specific discrepancies remain; most pro\-mi\-nent are:
\begin{itemize}
\item 
  the conspicuously low position as well as the decay properties of the
  negative parity $\Lambda_{\frac{1}{2}^-}(1405)$ resonance; The calculated mass
  of this state exceeds the experimental value by more than 100 MeV;
\item
  there is experimental evidence~\cite{PDG} for excited negative parity
  $\Delta$-resonances well below 2 GeV which can not be accounted for by the
  quark model mentioned above, see fig.~\ref{Intro_Fig1},
  \begin{figure}[!htb]
    \centering
    \includegraphics[width=\linewidth]{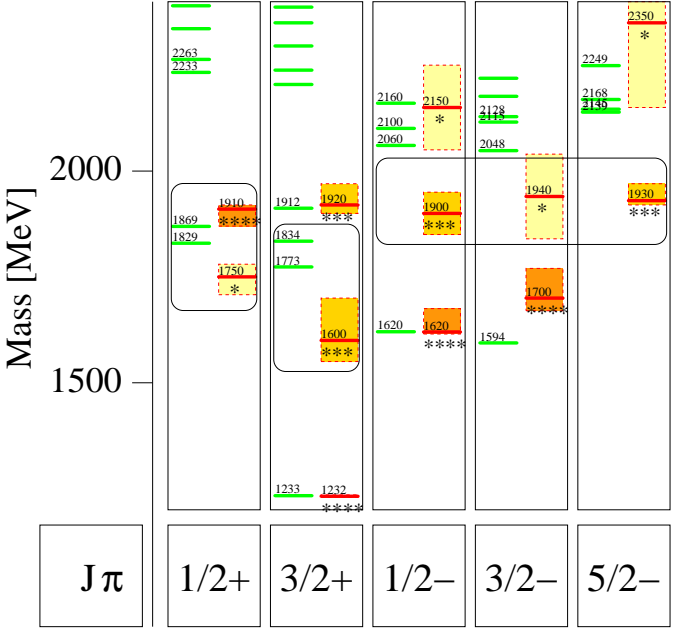}
    \caption{
      Discrepancies in the $\Delta$ mass spectrum: The left part of each
      column represents the results obtained in model $\mathcal{A}$
      of~\cite{LoeMePe2} in comparison with experimental data from the Particle Data
      Group~\cite{PDG} (right side of each column), where lines are the resonance
      position (mass) with the mass uncertainty represented by a shaded box and the
      rating of~\cite{PDG} indicated by stars. $J$ and $\pi$ denote total angular
      momentum and parity, respectively. Small differences with respect to the
      results from fig. 3 of~\cite{LoeMePe2} are due to the fact that we obtained
      increased numerical accuracy by diagonalising the resulting Salpeter
      Hamiltonian in larger model spaces, see section~\ref{subsec:BSAppr} for
      details.\label{Intro_Fig1}}
  \end{figure}
  nor by any other constituent quark model we are aware of;
\item
  The mass of the positive parity $\Delta_{\frac{3}{2}^+}(1600)$ resonance,
  see also fig.~\ref{Intro_Fig1}, the low value of which with respect to other
  excited states of this kind can not be traced back to instan\-ton-in\-duced
  effects, since these are absent for flavour symmetric states.
\end{itemize}

We therefore want to explore whether these deficiencies are inherent to the
constituent quark model itself or can be overcome by the introduction of an
additional quark interaction which improves upon the issues mentioned above
without deteriorating the excellent description of the majority of the other
states. In view of the fact that the discrepancies mainly affect the
$\Delta$-spectrum, this additional interaction is likely to be flavour
dependent. An obvious candidate in this respect would be a single pseudoscalar
meson exchange potential as has been used as a basis of an effective
spin-flavour dependent quark interaction very successfully by the Graz-group
\cite{Glozman1996,Glozman1997,Glozman1998_1,%
Glozman1998_2,Theussl,Glantschnig,Melde08,Plessas}.

In the present paper we shall investigate various implementations of the
coordinate (or momentum) dependence of such interactions. The paper is
organised as follows: After a brief recapitulation of the ingredients and
basic equations of our Bethe-Salpeter model (for more details
see~\cite{LoeMePe1}) in section~\ref{sec:BSmodel}, we discuss in
section~\ref{sec:BSint} the form and the parameters of the effective quark
interactions used in this paper. Section~\ref{sec:BSres} contains the results
and a discussion of the baryon mass spectra in comparison to the results
obtained before~\cite{LoeMePe2,LoeMePe3}. In
section~\ref{sec:electromagneticFF} we present some results on ground state
form factors before concluding in section~\ref{sec:Conc}.

\section{Bethe-Salpeter model}
\label{sec:BSmodel}

\subsection{Bound state Bethe-Salpeter amplitudes}
\label{subsec:BSBS}
The basic quantity describing three-quark bound states is the Bethe-Salpeter
amplitude $\chi$ defined in position space through
\begin{eqnarray}
  \label{eq:BSmodel1}
  \lefteqn{
    \chi_{\bar{P}\,a_1 a_2 a_3}(x_1,x_2,x_3)
  }
  \nonumber\\
  &&=
  \langle 0| 
  T\,\Psi_{a_1}(x_1)\Psi_{a_2}(x_2)\Psi_{a_3}(x_3)
  |\bar P\rangle\,,
\end{eqnarray}
where $T$ is the time ordering operator, $\left|\bar P\right\rangle$
represents the bound-state with total 4 momentum $\bar P^2=M^2$ of a baryon
with mass $M$\,, $\left|0\right\rangle$ is the physical vacuum and
$\Psi_{a_i}(x_i)$ denotes single quark-field operators with multi-indices
$a_i$ in Dirac, colour and flavour space. Because of translational invariance
below we shall exclusively use relative Jacobi coordinates $p_\xi, p_\eta$ in
momentum space. The Fourier transform of the Bethe-Salpeter amplitudes:
$\chi_{\bar P\,a_1a_2a_3}(p_{\xi},p_{\eta})$ are then determined by the
homogeneous \\
Bethe-Salpeter equation compactly written as
\begin{equation}
  \label{eq:BSmodel2}
  \chi_{\bar P}
  =
  -\textrm{i}\,
  G_{0\,,P} \left(
    K^{(3)}_{\bar P} 
    + 
    \bar{K}^{(2)}_{\bar P}
  \right)\,
  \chi_{\bar P}\,,
\end{equation}
where $K^{(3)}_P$ represents the irreducible 3-quark-kernel and where
$\bar{K}^{(2)}_P$ is defined by
\begin{eqnarray}
  \label{eq:BSmodel3}
  \lefteqn{
    \bar{K}^{(2)}_P(p_{\xi},p_{\eta};p_{\xi}',p_{\eta}') 
    =
    \sum_{k=1}^3
    (2\pi)^4\,\delta^{(4)}(p_{\eta_k} - p_{\eta_k}')\,
  }
  \nonumber\\
  && 
  \times K^{(2)}_{(\frac{2}{3}P+p_{\eta_k})}
  (p_{\xi_k},p_{\xi_k}')
  \otimes 
  \left(S_F^3\right)^{-1}\!\!\left(\textstyle\frac{P}{3}-p_{\eta_k}\right)
\end{eqnarray}
in terms of the irreducible two-body interaction kernel 
\\
$K^{(2)}_{\!(\!\frac{2}{3}P+p_{\eta_k}\!)}$ for each quark pair labeled by the
odd-particle index $k$. 
Furthermore $G_{0,P}$ is the free 3-quark fermion propagator defined as
\begin{eqnarray}
  \label{eq:BSmodel4}
  \lefteqn{G_{0,P}(p_{\xi},p_{\eta};p_{\xi}',p_{\eta}') =}
  \nonumber\\
  &&
  (2\pi)^8\,\delta^{(4)}(p_{\xi}-p_{\xi}')\,\delta^{(4)}(p_{\eta}-p_{\eta}')
  \nonumber\\
  &&
  \times S^1_F(\textstyle\frac{P}{3}\!+\!p_{\xi}\!+\!\frac{p_{\eta}}{2}) 
  \!\otimes\!
  S^2_F(\frac{P}{3}\!-\!p_{\xi}\!+\!\frac{p_{\eta}}{2})
  \!\otimes\!
  S^3_F(\frac{P}{3}\!-\!p_{\eta})
\end{eqnarray}
in terms of full single quark propagators $S^i_F$\,.

\subsection{Model assumptions}
\label{subsec:BSAppr}
In view of the fact that the interaction kernels and the propagators are sums
of infinitely many Feynman diagrams, in order to arrive at a tractable model
we make the following assumptions, mainly with the goal to stay in close
contact with the quite successful non-relativistic constituent quark model:
\begin{itemize}
\item 
  The full propagators $S^i_F$ are replaced by Feynman
  pro\-pagators of the free form
  \begin{equation}
    \label{eq:BSmodel5}
    S^i_F(p) 
    \overset{!}{=} 
    \frac{\textrm{i}}{p\dslash - m_i + \textrm{i}\,\varepsilon}\,,
  \end{equation}
  tacitly assuming that at least some part of the self-energy can effectively
  be subsumed in an effective constituent quark mass $m_i$ which then is a 
  parameter of the model. 
\item
  Obviously this does not account for confinement: This is assumed to be
  implemented in the form of an instantaneous interaction kernel which in the
  rest frame of the baryon is described by an unretarded potential $V^{(3)}$\,:
  \begin{align}
    \label{eq:BSmodel6}
    K^{(3)}_P\left(p_\xi,p_\eta;{p_\xi}',{p_\eta}'\right) &
    \Big|_{P=(M,\vec 0)}
    \nonumber\\
    \overset{!}{=} \,& 
    V^{(3)}\left(\vec p_\xi,\vec p_\eta; {\vec p_\xi}',{\vec p_\eta}'\right)\,.
  \end{align}
  Likewise we assume that two-quark interaction kernels in the
  rest frame of the baryon are described by
  2-body potentials $V^{(2)}$\,:
  \begin{align}
    \label{eq:BSmodel7}
     K^{(2)}_{\frac{2}{3}P+p_{\eta_k}}\left(p_{\xi_k};{p_{\xi_k}}'\right) &
    \Big|_{P=(M,\vec 0)}
    \nonumber\\
    \overset{!}{=} \,&
    V^{(2)}\left(\vec p_{\xi_k}; {\vec p_{\xi_k}}'\right)\,.
  \end{align}
\end{itemize}
Then, with a perturbative elimination of retardation effects, which arise due
to the genuine two-body interactions, see ~\cite{LoeMePe1} for details, one
can derive an equation for the (projected) Salpeter amplitude
\begin{equation}
  \label{eq:BSmodel8}
  \Phi^\Lambda_M(\vec p_\xi,\vec p_\eta)
  =
  \Lambda_+(\vec p_\xi,p_\eta)
  \int
  \frac{\textrm{d}p^0_\xi}{2\pi}
  \frac{\textrm{d}p^0_\eta}{2\pi}
  \chi_M(p_\xi,p_\eta)\,,
\end{equation}
where
\begin{eqnarray}
  \label{eq:BSmodel9}
  \Lambda_{\pm}(\vec p_\xi,p_\eta)
  :=
  \Lambda^+(\vec p_1) 
  \otimes  
  \Lambda^+(\vec p_2) 
  \otimes  
  \Lambda^+(\vec p_3)
  \nonumber\\
  \pm\,\,
  \Lambda^-(\vec p_1) 
  \otimes  
  \Lambda^-(\vec p_2) 
  \otimes  
  \Lambda^-(\vec p_3)
\end{eqnarray}
with 
\begin{equation}
  \label{eq:BSmodel10}
  \Lambda^{\pm}(\vec p) 
  = 
  \sum_f \Lambda^{\pm}_{m_f}(\vec p) \otimes \mathcal{P}_f\,.
\end{equation}
Here
\begin{equation}
  \label{eq:BSmodel11}
  \Lambda^\pm_m(\vec p) := \frac{\omega_m(\vec p) \pm H_m(\vec
    p)}{2\,\omega_m(\vec p)}
\end{equation}
are projection operators on positive and negative energy states and
$\mathcal{P}_f$ projects on quark flavour $f$\,. Furthermore 
the quark energy is given by $\omega(\vec p)=\sqrt{|\vec p|^2 + m^2}$ and
\begin{equation}
  \label{eq:BSmodel12}
  H_m(\vec p) := \gamma^0\left(\vec \gamma \cdot \vec p + m\right)
\end{equation}
is the Dirac Hamilton operator. As shown in detail in~\cite{LoeMePe1} this
equation can be written in the form of an eigenvalue problem
\begin{equation}
  \label{eq:BSmodel13}
  \mathcal{H}\,\Phi^\Lambda_M = M\,\Phi^\Lambda_m
\end{equation}
for the projected Salpeter amplitude, where the eigenvalues are the baryon
masses $M$\,. The Salpeter Hamilton operator is given by
\begin{eqnarray}
  \label{eq:BSmodel14}
  \lefteqn{
    \left[\mathcal{H}\,\Phi^\Lambda_M\right](\vec p_\xi,\vec p_\eta)
    =
    \mathcal{H}_0(\vec p_\xi,\vec p_\eta)\,\Phi^\Lambda_M(\vec p_\xi,\vec p_\eta)
  }
  \nonumber\\
  &+&
  \Lambda_+(\vec p_\xi,\vec p_\eta)\,
  \gamma^0\!\otimes\!\gamma^0\!\otimes\!\gamma^0\!\otimes\!
  \int
  \frac{\textrm{d}^3{p_\xi}'}{(2\pi)^3}
  \frac{\textrm{d}^3{p_\eta}'}{(2\pi)^3}
  \nonumber\\
  &&
  \hspace*{3em}
  V^{(3)}(\vec p_\xi,\vec p_\eta; {\vec p_\xi}',{\vec p_\eta}')
  \,\Phi^\Lambda_M({\vec p_\xi}',{\vec p_\eta}')
  \nonumber\\
  &+&
  \Lambda_-(\vec p_\xi,\vec p_\eta)\,
  \gamma^0\!\otimes\!\gamma^0\!\otimes\!\mathds{1}
  \int
  \frac{\textrm{d}^3{p_\xi}'}{(2\pi)^3}
  \nonumber\\
  &&
  \hspace*{3em}
  \,V^{(2)}(\vec p_\xi; {\vec p_\xi}')\otimes\!\mathds{1}
  \,\Phi^\Lambda_M({\vec p_\xi}',{\vec p_\eta})
  \\
  &+&\textrm{corresponding quark interations (23) and (31)}\,\nonumber,
\end{eqnarray}
where $\mathcal{H}_0$ denotes the free three-quark Hamilton operator as a sum
of the corresponding single particle Dirac Hamilton operators.

The eigenvalue problem of Eq.~(\ref{eq:BSmodel13}) is solved by numerical
diagonalisation in a large but finite basis of oscillator states up to an
oscillator quantum number $N_{\textrm{max}}$\,. In previous
calculations~\cite{LoeMePe2,LoeMePe3} at least $N_{\textrm{max}}=12$ was
used. All the results in the present paper were obtained with at least
$N_{\textrm{max}}=18$\,, which, although computer time consuming, has the
advantage that for all states the independence of the numerical results on the
oscillator functions length scale in some scaling window could be warranted
and that all could be calculated with a universal value for this length
scale. This is a technical advantage when calculating electroweak amplitudes.

\section{Model Interactions}
\label{sec:BSint}
Below we specify the interaction potentials $V^{(3)}$ and $V^{(2)}$ used in
Eq.~(\ref{eq:BSmodel14})\,. These include a confinement potential, the
instanton induced two-quark interaction as has been used
before~\cite{LoeMePe2,LoeMePe3} and the new phenomenological potential
inspired by pseudoscalar meson exchange.

\subsection{Confinement}
\label{subsec:Conf}

Confinement is implemented by subjecting the quarks to a potential which rises
linearly with interquark distances, supplemented by an appropriate three
particle Dirac\linebreak structure $\Gamma$. The potential contains two parameters: the
off-set $a$ and the slope $b$ and is assumed to be of the following form in
coordinate space
\begin{equation}
  \label{eq:conf}
  V^{(3)}_{\textrm{conf}}(\vec x_1,\vec x_2,\vec x_3)\,
  =
  3\,a\,\Gamma_o
  + b \sum_{i<j} \left|\vec x_i -\vec x_j\right|\,\Gamma_s
\end{equation}
where $\Gamma_o$ and $\Gamma_s$ are suitably chosen Dirac
structures. Alternatively we can consider the linear potential to be treated
as a two-body kernel as will be used below.

\subsection{Instanton induced interaction}

Instanton effects leads to an effective quark-quark interaction, which for
quark pairs in baryons can be written in coordinate space as
\begin{eqnarray}
  \label{eq:III}
  \lefteqn{
    \mathcal{V}^{(2)}_{\textrm{III}}(x_1,x_2;x_1'x_2') 
    = 
    V^{(2)}_{\textrm{III}}(\vec x_1 - \vec x_2)
  }
  \nonumber\\
  &&\hspace*{2.5em}
  \times\,
  \delta(x_1^0-x_2^0)\,
  \delta^{(4)}(x_1-x_1')\,
  \delta^{(4)}(x_2-x_2')
\end{eqnarray}
with
\begin{eqnarray}
  \label{eq:III2}
    V^{(2)}_{\textrm{III}}(\vec x)\,
  &=&
  -4v(\vec x)
  \left(
    \mathds{1} \!\otimes\! \mathds{1} + \gamma^5 \!\otimes\! \gamma^5
  \right)
  \mathcal{P}^\mathcal{D}_{S_{12}=0}
  \nonumber\\
  &&
  \hspace*{-2.5em}\!\otimes\!
  \left(
    g_{nn}\,\mathcal{P}^\mathcal{F}_\mathcal{A}(nn)
    +
    g_{ns}\,\mathcal{P}^\mathcal{F}_\mathcal{A}(ns)
  \right)\,,
\end{eqnarray}
where $\mathcal{P}^\mathcal{D}_{S_{12}=0}$ is a projector on spin-singlet
states and $\mathcal{P}^{\mathcal{F}}_\mathcal{A}(f_1 f_2)$ projects
on flavour-antisymmetric quark pairs with fla\-vours $f_1$ and $f_2$.
Although the two couplings $g_{nn}$ and $g_{ns}$ are in principle determined
by integrals over instanton densities, these are treated as free parameters
here. As it stands this is a contact interaction, which for our purpose is
regularised by replacing the coordinate space dependence by a Gaussian
\begin{equation}
  \label{eq:III3}
  v_\lambda(\vec x) 
  = 
  \frac{1}{\lambda^3\,\pi^{\frac{3}{2}}}\,
  \exp\left(-\tfrac{|\vec x|^2}{\lambda^2}\right)\,.  
\end{equation}
The effective range parameter $\lambda$ is assumed to be flavour independent
and enters as an additional parameter.

\subsection{An additional flavour dependent interaction}
\label{subsec:Afdi}

The coupling of spin-$\frac{1}{2}$ fermions to a flavour nonet of 
pseudoscalar meson fields is
given by an interaction Lagrange density
\begin{equation}
  \label{eq:OBEps}
  \mathcal{L}_I^{\textnormal{(ps)}} 
  =
  -\mathrm{i}
  \sum_{a=0}^8
  g_a\,\bar{\psi}\,\gamma^5\,\lambda^a\,\psi\,\phi^a,
\end{equation}
in the case of so-called pseudoscalar coupling and by
\begin{equation}
  \label{eq:OBEpv}
  \mathcal{L}_I^{\textnormal{(pv)}} 
  =
  -\sum_{a=0}^8
  \frac{g_a}{2m}\,\bar{\psi}\,\gamma^5\,\gamma^{\mu}\,
  \lambda^a\,\psi\,\partial_{\mu}\phi^a.
\end{equation}
for pseudovector coupling. Here $\psi$ represents the quark fields
with mass $m$ and $\phi^a$ the pseudoscalar meson fields with mass $\mu_a$
where the flavour index $a=\pi^{\pm\,,0},\eta^0_8,$ $\eta_1^0$ $K^\pm,$ $K^0,$
$\bar K^0$\,. The flavour dependence is represented by the usu\-al Gell-Mann
matrices $\lambda^a, a=1,\ldots,8$\,; $\lambda^0$ is proportional to the
identity operator in flavour space normalised to 
$\textrm{Tr}((\lambda^0)^2)=2$.

A standard application of the Feynman rules, see \emph{e.g.} ~\cite{CaDuElHaSiSp} 
then leads to the second order scattering-matrix element
$\mathcal{M}^{(2)}$ given in the CM-system by the expressions
\begin{eqnarray}
  \label{eq:OBEM2ps}
  \lefteqn{
    \mathrm{i}
    \mathcal{M}_{\textnormal{(ps)}}^{(2)}(k_0,\vec{k})
  }
  \nonumber\\
  &=&
  \sum_{a,b} g_a^2
  \bigg[
  \bar{\psi}(\vec{p}')(-i\gamma^5)\lambda^a\,\psi(\vec{p})\,
  D^{ab}(k_0,\vec{k})
  \nonumber\\
  &&
  \hspace*{1em}\times\bar{\psi}(-\vec{p}')
  (-\mathrm{i}\gamma^5)
  \lambda^b
  \psi(-\vec{p})
  \bigg]
  \nonumber\\
  &=:& 
  -
  \sum_{a,b} g_a^2
  D^{ab}(k_0,\vec{k})
  \left[\lambda^a\,\gamma^5\right]
  \! \otimes\! 
  \left[\lambda^b\,\gamma^5\right]
  \nonumber\\
  &=:&
  \mathrm{i}
  [\bar{\psi}(-\vec{p}')\!\otimes\!\bar{\psi}(\vec{p}')]
  V_{(ps)}(k_0,\vec{k})
  [\psi(-\vec{p})\!\otimes\!\psi(\vec{p})] 
\end{eqnarray}
and
\begin{eqnarray}
  \label{eq:OBEM2pv}
  \lefteqn{
    \mathrm{i}\,\mathcal{M}_{\textnormal{(pv)}}^{(2)}(k_0,\vec{k})
  }
  \nonumber\\
  &=&
  \sum_{a,b} \frac{g_a^2}{4m^2}
  \bigg[
  \bar{\psi}(\vec{p}')\gamma^5\gamma^{\mu}(-ik_{\mu})
  \lambda_a\psi(\vec{p})\,
  D^{ab}(k_0,\vec{k})
  \nonumber\\
  &&
  \hspace*{1em}\times
  \bar{\psi}(-\vec{p}')\gamma^5\gamma^{\nu}(-i(-k_{\nu}))
  \lambda_b\psi(-\vec{p})
  \bigg]
  \nonumber\\
  &=:& 
  \sum_{a,b} 
  \frac{g_a^2}{4m^2}\,
  D^{ab}(k_0,\vec{k})\,k_{\mu}k_{\nu}\!
  \left[\lambda^a\gamma^5\gamma^{\mu}\right]\!\!\otimes\!\!
  \left[\lambda^b\gamma^5\gamma^{\nu}\right] 
  \nonumber\\
  &:=& 
  \mathrm{i}\,
  [\bar{\psi}(-\vec{p}')\!\otimes\!\bar{\psi}(\vec{p}')]
  V_{(pv)}(k_0,\vec{k})
  [\psi(-\vec{p})\!\otimes\!\psi(\vec{p})] 
\end{eqnarray}
in case of pseudoscalar and pseudovector coupling, respectively. Here the
meson propagator is given by
\begin{equation}
\label{eq:prop}
D^{ab}(k_0,\vec{k})
= 
\frac{\mathrm{i}\delta_{ab}}{k_0^2-|\vec{k}|^2-\mu_a^2}\,,
\end{equation}
with $k_{\mu}:=p_{\mu}'-p_{\mu}$ the momentum transfer.

In instantaneous approximation we set $k_0 = 0$. From
Eqs. [\ref{eq:OBEM2ps},{\ref{eq:OBEM2pv}] we extract the corresponding
potentials in momentum space
\begin{eqnarray}
  \label{eq:OBEVps}
  V_{(ps)}^{(2)}(\vec{k})
  =
  \sum_a g_a^2
  [\lambda^a\!\otimes\!\lambda^a]
  \frac{1}{|\vec{k}|^2+\mu_a^2}
  \bigg[\gamma^5\!\otimes\!\gamma^5\bigg]
\end{eqnarray}
for pseudoscalar coupling and
\begin{eqnarray}
  \label{eq:OBEVpv}
  \lefteqn{
    V_{(pv)}^{(2)}(\vec{k})
  }
  \\
  &=&
  \sum_a \frac{g_a^2}{4m^2}
  [\lambda^a\!\otimes\!\lambda^a]
  \frac{-1}{|\vec{k}|^2+\mu_a^2}
  \Big[\!
    \big(\gamma^5\vec{\gamma}\cdot\vec{k}\big)
    \!\otimes\!
    \big(\gamma^5\boldsymbol{\gamma}\cdot\vec{k}\big)
  \!\Big]
  \nonumber\\
  &=& 
  \sum_a 
  \frac{g_a^2}{4m^2}
  [\lambda^a\!\otimes\!\lambda^a]
  \frac{-|\vec{k}|^2}{|\vec{k}|^2+\mu_a^2}
  \Big[\!
    \big(\gamma^5\vec{\gamma}\cdot\widehat{\vec{k}}\big)
    \!\otimes\!
    \big(\gamma^5\boldsymbol{\gamma}\cdot\widehat{\vec{k}}\big)
  \!\Big]
  \nonumber
\end{eqnarray}
for pseudovector coupling, where $\widehat{\vec k} := \frac{\vec k}{|\vec
  k|}$\,.

As it stands, the expression for the potential in the instantaneous
approximation for pseudoscalar coupling\newline leads, after Fourier transformation,
to a local Yukawa potential in configuration space with the usual range given
by the mass of the exchanged pseudoscalar meson. For pseudovector coupling the
non-relativistic approximation to the Fourier transform leads to the usual
spin-spin contact interaction together with the usual tensor force. In the
simplest form adopted by the Graz group
\cite{Glozman1996,Glozman1997,Glozman1998_1,%
Theussl,Glantschnig,Melde08,Plessas}
the latter were ignored, in addition the contact term was regularised by a
Gaussian function and the Yukawa terms were regularised to avoid singularities
at the origin.

In view of this and the instantaneous approximation we decided to parametrise
the new flavour dependent interaction purely phenomenologically as a local
potential in configuration space, its simple form given by
\begin{equation}
  \label{eq:OBEVreg}
  V^{(2)}(\vec x)
  =
  \sum_a g_a^2\,
  \left[
    \lambda^a\,\gamma^5 \!\otimes\! \lambda^a\,\gamma^5
  \right]\,
  v_{\lambda_a}(\vec x)
\end{equation}
where $v_\lambda(\vec x)$ is the Gaussian form given in Eq.~\ref{eq:III3}\,.
Other Dirac structures, such as 
$\left[\gamma^5 \vec \gamma\cdot \widehat{\vec x}\,
  \!\otimes\! 
  \gamma^5 \vec \gamma\cdot \widehat{\vec x}\right]$
were tried, but were found to be less effective. Results for the meson
exchange form of the interaction as given by
Eqs.~(\ref{eq:OBEVps},\ref{eq:OBEVpv}) will be briefly discussed in
section~\ref{sec:Conc}.

\begin{table}[!htb]
\centering
\caption{
The list of baryon resonances of which the masses were used to determine the
model parameters in a least-squares fit where every resonance was attributed a
weight reciprocal to its uncertainty in its position as given in \cite{PDG}.
Nominal masses are given in MeV.
\label{tab:Tab_3}
}
\begin{tabular}{
@{\hspace{0pt}}l@{\hspace{2pt}}l@{\hspace{2pt}}l@{\hspace{2pt}}%
l@{\hspace{6pt}}l@{\hspace{0pt}}%
}
\toprule
$\Delta$ & $N$ & $\Lambda / \Sigma$ & $\Xi$ & $\Omega$ \\
\midrule
$S_{31}(1620)$   & $S_{11}(1535)$ & $S_{21}(1620)$ & $S_{11}(1309)$ & \\
$S_{31}(1900)$   & $S_{11}(1650)$ & $S_{21}(1750)$ & & \\
\midrule
$P_{31}(1750)$   & $P_{11}(939)$  & $P_{01}(1116)$ & $P_{13}(1530)$ & $P_{01}(1672)$\\
$P_{31}(1910)$   & $P_{11}(1440)$ & $P_{21}(1289)$ & & \\
$P_{33}(1232)$   &               & $P_{01}(1600)$ & &  \\
$P_{33}(1600)$   &               & $P_{21}(1660)$ & & \\
$P_{33}(1920)$   &               &               & & \\
\midrule
$D_{33}(1700)$   &               &               & $D_{13}(1820)$ & \\
$D_{33}(1940)$   &               &               & & \\
$D_{35}(1930)$   &               &               & & \\
\midrule
$F_{35}(1905)$   &               &               & & \\
$F_{35}(2000)$   &               &               & & \\
$F_{37}(1950)$   &               &               & & \\
\midrule
$G_{37}(2200)$   &               &               & & \\
$G_{39}(2400)$   &               &               & & \\
\midrule
$H_{39}(2300)$   &               &               & & \\
$H_{3\,11}(2420)\!\!\!$ &        &                & & \\
\midrule
$K_{3\,15}(2950)\!\!\!$ &        &                & & \\
\bottomrule
\end{tabular}
\end{table}

\section{Mass spectra}
\label{sec:BSres}

\subsection{Model parameters}
\label{subsec:parameters}
The resulting baryon mass spectra were obtained by fitting the parameters of
the model, \emph{viz.} the offset $a$ and slope $b$ of the confinement
potential, the constituent quark masses $m_n=m_u=m_d$ and $m_s$, the strengths
of the instanton induced force, $g_{nn}$ and $g_{ns}$ as well as the strengths
of the additional flavour dependent interaction, given by $g_8$ and $g_0$ for
flavour octet and flavour singlet exchange (thus assuming $SU(3)$ symmetry) to
a selection of baryon resonances, see table~\ref{tab:Tab_3}\,.
The range $\lambda$ given to the instanton induced force was kept
to the value used in~\cite{LoeMePe2,LoeMePe3} and is roughly in accordance
with typical instanton sizes. The optimal value for the range of the
additional flavour dependent interaction was found to be $\lambda_8 = \lambda_0
\approx 0.25\,\textnormal{fm}$ and thus turned out to be of rather short range.
A comparison of the parameters obtained with the parameters of model
$\mathcal{A}$ of~\cite{LoeMePe2,LoeMePe3} is given in table~\ref{tab:par}\,.
\begin{table}[!htb]
\centering
\caption{Model parameters for the current model $\mathcal{C}$ in
  comparison to those of model $\mathcal{A}$ of~\cite{LoeMePe2,LoeMePe3}.
  Some of the parameters have been slightly changed with respect to
  the original values (listed in brackets) of~\cite{LoeMePe2,LoeMePe3},
  since the calculation has been performed with higher numerical accuracy
  by taking more basis states in the diagonalisation of the Salpeter
  Hamiltonian, see also text.
 \label{tab:par}}
\begin{tabular}{l@{\hspace{10pt}}l@{\hspace{10pt}}r@{\hspace{10pt}}r}
\toprule
parameter                    &                                        & model $\mathcal{C}$      & model $\mathcal{A}$ \\
\midrule
masses                       & $m_n$ [MeV]                            &  325.0                   & 330.0 \\
                             & $m_s$ [MeV]                            &  600.0                   & 670.0 \\
\midrule
\multirow{2}{*}{confinement} & \multirow{2}{*}{$a$ [MeV]}             & \multirow{2}{*}{-366.78} & -734.6 \\
                             &                                        &                          & [-744.0] \\
                             & \multirow{2}{*}{$b$ [MeV/fm]}          &  \multirow{2}{*}{212.81} &  453.6\\
                             &                                        &                          & [440.0] \\
\midrule
\multirow{2}{*}{instanton}   & \multirow{2}{*}{$g_{nn}$ [MeV fm$^3$]} &  \multirow{2}{*}{341.49} & 130.3 \\
                             &                                        &                          & [136.0] \\
\multirow{2}{*}{induced}     & \multirow{2}{*}{$g_{ns}$ [MeV fm$^3$]} &  \multirow{2}{*}{273.55} & 81.8 \\
                             &                                        &                          & [96.0] \\
interaction                  & $\lambda$ [fm]                         &  0.4                     & 0.4 \\
\midrule
octet                        & $\frac{g_{8}^2}{4\pi}$  [MeV fm$^3$]   &  100.86                  & -- \\
exchange                     &                                        &                          & \\
singlet                      & $\frac{g_{0}^2}{4\pi}$ [MeV fm$^3$]    &  1897.43                 & -- \\
exchange                     &                                        &                          & \\
                             & $\lambda_{8}=\lambda_0$ [fm]           &  0.25                    & -- \\
\bottomrule
\end{tabular}
\end{table}
\begin{table}[!htb]
  \centering
  \caption{
    Multiplet decomposition of amplitudes of negative parity
    $\Delta$-resonances. For each amplitude the contribution to the Salpeter norm,
    see~\cite{LoeMePe1} is given in $\%$\,, in each row the upper line and the
    lower line give the positive and negative energy contribution,
    respectively. States are labeled by the calculated mass and $J^\pi$ denotes
    total angular momentum and parity, $^{2S+1}\mathcal{F}_{J}[D]$ label
    amplitudes with spin $S$, flavour representation with dimension $\mathcal{F}$,
    $SU(6)$ representation with dimension $D$\,. The dominant contribution is
    underlined.  \label{tab:LSD}
   }
  \begin{tabular}[c]{cc@{\hspace{30pt}}c@{\hspace{30pt}}rr}
    \toprule
    $J^\pi$ & Mass & pos. & $^{4}10_{J}[56]$ & $^{2}10_{J}[70]$ \\ 
            &      & neg. & $^{4}10_{J}[56]$ & $^{2}10_{J}[70]$ \\ 
    \midrule        
    $\frac{1}{2}^-$ & 1635 & 98.9 & 6.6 & \underline{92.3} \\
                    &      &  1.1 & 0.8 & 0.4 \\
    \midrule              
    $\frac{1}{2}^-$ & 1932 & 98.5 & 14.2 & \underline{84.3} \\
                    &      &  1.5 & 1.0 & 0.5 \\
    \midrule              
    $\frac{1}{2}^-$ & 2041 & 99.0 & \underline{81.6} & 17.3 \\
                    &      &  1.0 & 0.5 & 0.5 \\
    \midrule              
    $\frac{3}{2}^-$ & 1592 & 98.2 & 9.9 & \underline{88.4} \\
                    &      &  1.8 & 0.9 & 0.9 \\
    \midrule              
    $\frac{3}{2}^-$ & 1871 & 97.9 & 24.5 & \underline{73.3} \\
                    &      &  2.1 & 1.4 & 0.7 \\
   \midrule              
   $\frac{3}{2}^-$ & 1944 & 98.6 & \underline{65.1} & 33.5 \\
                   &      &  1.4 & 0.9 & 0.6 \\
   \midrule              
   $\frac{5}{2}^-$ & 2013 & 98.8 & \underline{89.3} & 9.6 \\
                   &      &  1.2 & 0.6 & 0.6 \\
    \midrule              
    $\frac{5}{2}^-$ & 2136 & 99.2 & 16.8 & \underline{82.4} \\
                    &      &  0.8 & 0.3 & 0.5 \\
    \midrule              
    $\frac{5}{2}^-$ & 2166 & 99.1 & \underline{86.6} & 12.4 \\
                    &      &  1.0 & 0.4 & 0.6 \\
    \bottomrule
  \end{tabular}
\end{table}
The parameters need some comments: In the original
paper~\cite{LoeMePe2,LoeMePe3} the Dirac structures (\emph{i.e.} the spin
dependence) of the confinement potential were taken to be 
$\Gamma_0 = \frac{1}{4}(
\mathds{1}\!\otimes\!\mathds{1}\!\otimes\!\mathds{1}
+\gamma^0\!\otimes\!\gamma^0\!\otimes\!\mathds{1}
+\textrm{cycl. perm.})$
and 
$
\Gamma_s = \frac{1}{2}(
-\mathds{1}\!\otimes\!\mathds{1}\!\otimes\!\mathds{1}
+\gamma^0\!\otimes\!\gamma^0\otimes\mathds{1}
+\textrm{cycl. perm.})$
for the offset and slope, respectively, and were considered to build a 3-body
kernel. In the present model including the additional octet and singlet
flavour exchange potential into account we obtained the best results with
$\Gamma_0 =
\mathds{1}\!\otimes\!\mathds{1}\!\otimes\!\mathds{1}
$
and 
$
\Gamma_s = 
\gamma^0\!\otimes\!\gamma^0
$
and treating the interaction corresponding to the latter term as a 2-body
interaction. This of course impedes a direct comparison of the corresponding
parameters. Furthermore it was found that the strengths of the instanton
induced interaction is roughly tripled when compared to the original
values. Note that the additional flavour exchange interaction has the same
spin-flavour dependence as parts of the former interaction. The flavour
singlet exchange could effectively be considered as an other spin dependent
part of the confinement potential. Possibly this explains the extraordinary
large coupling in this case. In summary it thus must be conceded that the
present treatment is phenomenological altogether and that here unfortunately
the relation to more fundamental QCD parameters, such as instanton couplings
and string tension is lost. Nevertheless with only 10 parameters we consider
the present treatment to be effective especially in view of its merits in the
improved description of some resonances to be discussed below.

\subsection{The $\Delta$ and the $\Omega$ spectrum}
\label{subsec:DeltaSpectrum}

In fig.~\ref{Fig_GaussDeltaPSMeson} we compare the results from the present
calculation (model $\mathcal{C}$) (right side of each column) with
experimental data from the Particle Data Group~\cite{PDG} (central in each
column) and with the results from model $\mathcal{A}$ of ~\cite{LoeMePe2}
(left side in each column)\,. The parameters used are listed in
table~\ref{tab:par}\,.
\begin{figure*}[!htb]
\centering
\includegraphics[width=\linewidth]{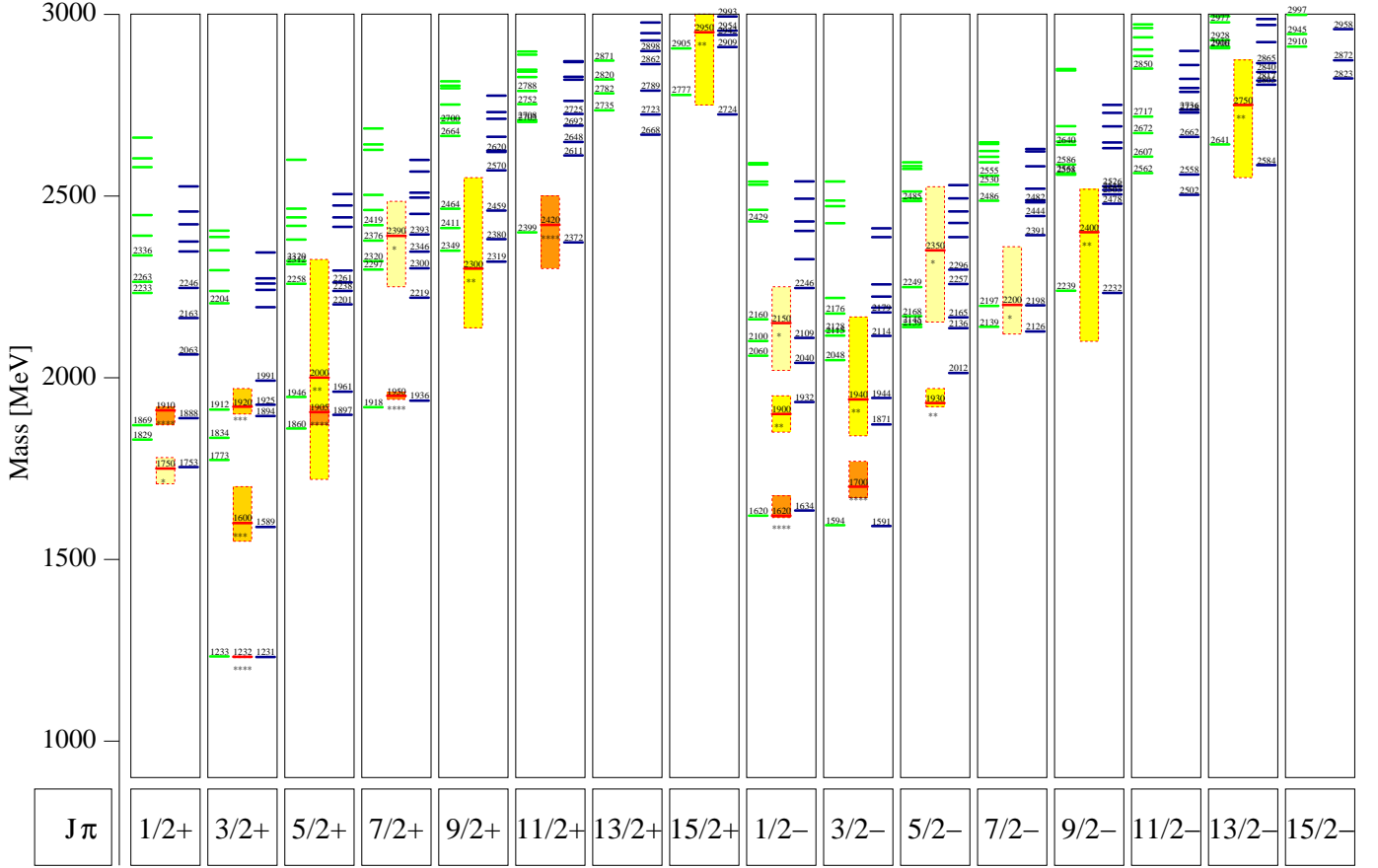}
\caption{
  Comparison of the $\Delta$-Spectrum calculated within the present model
  $\mathcal{C}$ (right side of each column) with experimental data from the
  Particle Data Group~\cite{PDG} (central in each column) and with the results
  from model $\mathcal{A}$ of ~\cite{LoeMePe2} (left side in each column), note
  the caption to fig.~\ref{Intro_Fig1}; Lines indicate the resonance position
  (mass) with the mass uncertainty represented by a shaded box and the rating
  of~\cite{PDG} indicated by stars. The small numbers give the mass in MeV. $J$
  and $\pi$ denote total angular momentum and parity, respectively.
  \label{Fig_GaussDeltaPSMeson}}
\end{figure*}
\begin{table}[!htb]
\centering
\caption{
  Comparison of experimental~\cite{PDG} and calculated masses in MeV of
  $\Delta$-resonances. The corresponding spectra 
  are shown in fig.~\ref{Fig_GaussDeltaPSMeson}\,.
  \label{tab:Delta}
}
\begin{tabular}{%
    *{6}{@{\hspace{2pt}}r}%
  }
  \toprule
  \multicolumn{1}{c}{exp.} & rating        & model $\mathcal{C}$ &
  \multicolumn{1}{c}{exp.} & rating        & model $\mathcal{C}$ \\
  \midrule
        $S_{31}(1620)$     & \tiny{****}   & 1634      & 
        $S_{31}(1900)$     & \tiny{***}    & 1932
        \\
        $S_{31}(2150)$     & \tiny{*}      & 2040/2109
                           &               & 
        \\
        \midrule
        $P_{31}(1750)$     & \tiny{*}      & 1653      & 
        $P_{31}(1910)$     & \tiny{****}   & 1888 
        \\
        $P_{33}(1232)$     & \tiny{****}   & 1231      &
        $P_{33}(1600)$     & \tiny{***}    & 1559
        \\
        $P_{33}(1920)$     & \tiny{***}    & 1894/1925
                           &               &
        \\
        \midrule
        $D_{33}(1700)$     & \tiny{****}   & 1591      & 
        $D_{33}(1940)$     & \tiny{*}      & 1871/1944
        \\
        $D_{35}(1930)$     & \tiny{***}    & 2012
                           &               &
        \\
        \midrule
        $F_{35}(1905)$     & \tiny{****}   & 1897      & 
        $F_{35}(2000)$     & \tiny{*}      & 1961
        \\
        $F_{37}(1950)$     & \tiny{****}   & 1936      & 
        $F_{37}(2390)$     & \tiny{*}      & many res.
        \\
        \midrule
        $G_{37}(2200)$     & \tiny{*}      & 2126/2198 & 
        $G_{39}(2400)$     & \tiny{**}     & 2232\\
        \midrule
        $H_{39}(2300)$     & \tiny{**}     & 2319      & 
        $H_{3,11}(2420)$   & \tiny{****}   & 2372
        \\
        \midrule
        $I_{3,13}(2750)$   & \tiny{**}     & 2584      & 
                           &               &
        \\
        \midrule
        $K_{3,15}(2950)$   & \tiny{**}    & 2724/2909  & 
                           &              &
        \\
        \bottomrule
\end{tabular}
\end{table}
\begin{table}[!htb]
\centering
\caption{%
  Comparison of experimental~\cite{PDG} and calculated masses in MeV of
  $N$-resonances. The corresponding spectra 
  are shown in fig.~\ref{Fig_GaussNucleonPSMeson}\,.
  \label{tab:Tabnine}
}
\begin{tabular}{%
*{6}{@{\hspace{2pt}}r}
}
  \toprule
  \multicolumn{1}{c}{exp.} & rating & model $\mathcal{C}$ &
  \multicolumn{1}{c}{exp.} & rating & model $\mathcal{C}$ \\
  \midrule
      $S_{11}(1535)$     & \tiny{****}                   & 1484 & 
      $S_{11}(1650)$     & \tiny{****}                   & 1672
      \\
      $S_{11}(2090)$     & \tiny{*}                    & many res.&
      & & 
      \\
      \midrule
      $P_{11}(939)$      & \tiny{****}                   &  945 & 
      $P_{11}(1440)$     & \tiny{****}                   & 1440
      \\
      $P_{11}(1710)$     & \tiny{***}                    & 1709 & 
      $P_{11}(2100)$     & \tiny{*}                    & many res.
      \\
      $P_{13}(1720)$     & \tiny{****}                   & 1703 & 
      $P_{13}(1900)$     & \tiny{**}                    & 1825\\
      \midrule
      $D_{13}(1520)$     & \tiny{****}                   & 1534 & 
      $D_{13}(1700)$     & \tiny{***}                    & 1685
      \\
      $D_{13}(2080)$     & \tiny{**}                    & many res.& 
      $D_{15}(1675)$     & \tiny{****}                   & 1667
      \\
      $D_{15}(2200)$     & \tiny{**}                    & many res.
      & & 
      \\
      \midrule
      $F_{15}(1680)$     & \tiny{****}                   & 1761 & 
      $F_{15}(2000)$     & \tiny{**}                    & 1930/1983
      \\
      $F_{17}(1990)$     & \tiny{**}                    & 2001 & 
      & & 
      \\
      \midrule
      $G_{17}(2190)$     & \tiny{****}                   & 2011 & 
      $G_{19}(2250)$     & \tiny{****}                   & 2165
      \\
      \midrule
      $H_{19}(2220)$     & \tiny{****}                   & 2192 & 
      & & 
      \\
      \midrule
      $I_{1,11}(2600)$ & \tiny{***}                    & 2377 & 
      & & 
      \\
      \midrule
      $K_{1,13}(2700)$ & \tiny{**}                    & 2543 & 
      & & 
      \\
      \bottomrule
\end{tabular}
\end{table}
The spectrum of the $\Delta$ (see fig.~\ref{Fig_GaussDeltaPSMeson}) and
$\Omega$ (see right panel of fig. \ref{Fig_GaussXiOmegaPSMeson}}) resonances
is determined by the confinement potential and the flavour exchange
interaction only, since the instanton induced interaction does not act on
flavour symmetric states. Concerning the positive parity resonances we see
that in the present calculation we can now indeed account for the low position
of the $\Delta_{\frac{3}{2}^+}(1600)$ resonance. In addition the next
excitations in this channel now lie closer to 2000 MeV in better agreement
with experimental data, as is also the case for the splitting of the two
$\Delta_{\frac{1}{2}^+}$ resonances.  Note, however, that these states were
included in the parameter fit.  Additionally there is support for a parity
doublet $\Delta_{\frac{3}{2}^+}(1920)$ and $\Delta_{\frac{3}{2}^-}(1940)$ as
argued by~\cite{Horn}.

Likewise, we can now account for the excited negative parity resonances:
$\Delta_{\frac{1}{2}^-}(1900)$\,,$\Delta_{\frac{3}{2}^-}(1940)$ and
$\Delta_{\frac{5}{2}^-}$ $(1930)$ and even find two states in the
$\Delta_{\frac{3}{2}^-}$ channel which could correspond to the poorly
established $\Delta_{\frac{3}{2}^-}(1940)$ state. 

In view of the near degeneracy of the $\Delta_{\frac{1}{2}^-}$ $(1900)$\,,
$\Delta_{\frac{3}{2}^-}$ $(1940)$ and $\Delta_{\frac{5}{2}^-}(1930)$ states it
is tempting to classify these in a non-relativistic scheme as a total spin
$S=\frac{3}{2}$, total quark angular momentum $L=1$ multiplet, which, because
of total isospin $I=\frac{3}{2}$ must then belong to a $(56,1^-)$ multiplet,
which is lowered with respect to the bulk of the other negative parity states
that in an oscillator classification would be attributed to the $N=3$ band.

Obviously this is not supported by the calculations: As table~\ref{tab:LSD}
shows, although the lowest $J^\pi =\frac{5}{2}^-$ resonance has a dominant
component in this multiplet, the second excited $J^\pi = \frac{1}{2}^-,
\frac{3}{2}^-$ resonances have dominant components in the $(70,1^-)$
multiplet; indeed the third excited states in these channels can be attributed
to the $(56,1^-)$ multiplet.  Otherwise the description and in particular the
$\Delta$-Regge-trajectory are of a similar quality as in the original
model $\mathcal{A}$\,.

Concerning the $\Omega$ spectrum, see fig.~\ref{Fig_GaussXiOmegaPSMeson}
(right panel), apart from the appearance of an excited
$\Omega_{\frac{3}{2}^+}$ state at 2021 MeV no spectacular changes in the
predictions with respect to the original model $\mathcal{A}$ were found. Note
that the present model predicts that this state is almost degenerate with the
first negative parity states $\Omega_{\frac{1}{2}^-}$ at 2008 MeV and
$\Omega_{\frac{3}{2}^-}$ at 1983 MeV.

In table~\ref{tab:Delta} we have summarised the calculation of
$\Delta$-res\-on\-an\-ces.

\subsection{The $N$ spectrum}
\label{subsec:NucleonSpectrum}

\begin{figure*}[!htb]
  \centering
  \includegraphics[width=\linewidth]{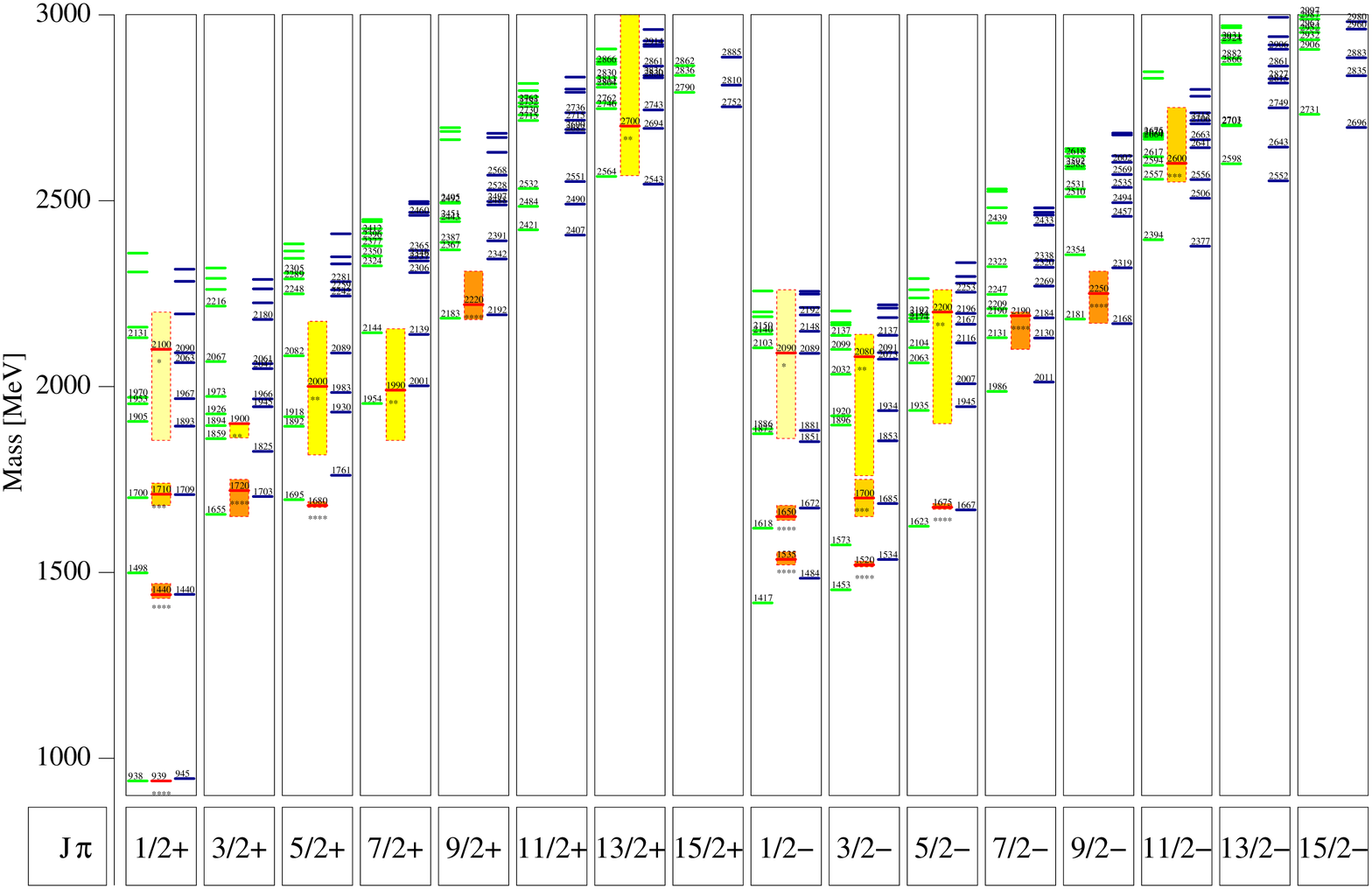}
  \caption{
    Comparison of the $N$-Spectrum calculated within the present model
    $\mathcal{C}$ (right side of each column) with experimental data from the
    Particle Data Group~\cite{PDG} (central in each column) and with the results
    from model $\mathcal{A}$ of ~\cite{LoeMePe2} (left side in each column).
    See also caption to fig.~\ref{Fig_GaussDeltaPSMeson}.
    \label{Fig_GaussNucleonPSMeson}
  }
\end{figure*}

In fig.~\ref{Fig_GaussNucleonPSMeson} we present the results for the nucleon
spectrum.
As was the case for the $\Delta$ spectrum, comparing to the results from the
former model $\mathcal{A}$ we indeed obtain an improved description of the
position of the first excited state with the same quantum numbers as the
ground state, the so called Roper resonance, while at the same time improving
also on the position of the first excited negative parity resonances $J^\pi =
\big(\tfrac{1}{2}^-\big)_{1}$,$\big(\tfrac{1}{2}^-\big)_{2}$,
$\big(\tfrac{3}{2}^-\big)_{1}$, $\big(\tfrac{3}{2}^-\big)_{2}$ and
$\big(\tfrac{5}{2}^-\big)_{1}$\,. With the exception of the
$J^\pi=\frac{5}{2}^+$ state, which compared to model $\mathcal{A}$ is shifted
upwards by approximately 50 MeV, the description of all known excited states
is of a similar quality as that of model $\mathcal{A}$. In particular the
position of the lowest $J^\pi = \frac{7}{2}^-$ resonance is still
underestimated by more than 100 MeV. 

In the following we compare the predictions obtained in model $\mathcal{C}$
for nucleon resonances with $J \le \frac{5}{2}$ and masses larger than 1.8 GeV
with new results obtained in the Bonn-Gatchina analyses as reported in
\cite{Anisovich_1,Anisovich_2}: In particular in \cite{Anisovich_1} a fourth
$J^\pi = \frac{1}{2}^+$ state was found, called $N_{\frac{1}{2}^+}(1875)$
which could correspond to our calculated state at 1893 MeV. Furthermore the
analysis contains two\linebreak $J^\pi = \frac{3}{2}^+$ states, called $N(1900)P_{13}$
and $N_{\frac{3}{2}^+}(1975)$, which might be identified with the model
$\mathcal{C}$ states calculated at 1825 MeV and 1945 MeV (or 1966 MeV),
respectively. Concerning the negative parity states, in \cite{Anisovich_2} a
new $J^\pi = \frac{1}{2}^-$ state was found ($N_{\frac{1}{2}^-}(1895)$) which
could be identified with the calculated state at 1851 MeV (or with that at
1881 MeV). In addition two $J^\pi = \frac{3}{2}^-$ states were found:
$N_{\frac{3}{2}^-}(1875)$ could correspond to the calculated states at 1853
MeV (or at 1934 MeV) and $N_{\frac{3}{2}^-}(2150)$ with one of the three
states with calculated masses 2073 MeV, 2091 MeV and 2137 MeV. The new
$N_{\frac{5}{2}^-}(2060)$ state reported in \cite{Anisovich_2} is closest to
the states calculated at 1945 MeV and at 2007 MeV. Finally, for $J^\pi =
\frac{5}{2}^+$ the analysis is ambiguous: Although a solution with a single
pole around 2.1 GeV is not excluded, solutions with 2 poles, either an
ill-defined pole in the 1800-1950 MeV mass region and one at nearly 2.2 GeV or
two close poles at approximately 2.0 GeV were found and could correspond to
the model $\mathcal{C}$ states calculated at 1930 MeV and 1983 MeV. Note that
these new resonances were not included in the parameter fit (see table~\ref{tab:Tab_3}).
An overview of the identification of nucleon resonances is given in
table~\ref{tab:Tabnine}.

\subsection{Hyperon spectra}
\label{subsec:YSpectra}
\begin{figure*}[!htb]
  \centering
  \includegraphics[width=\linewidth]{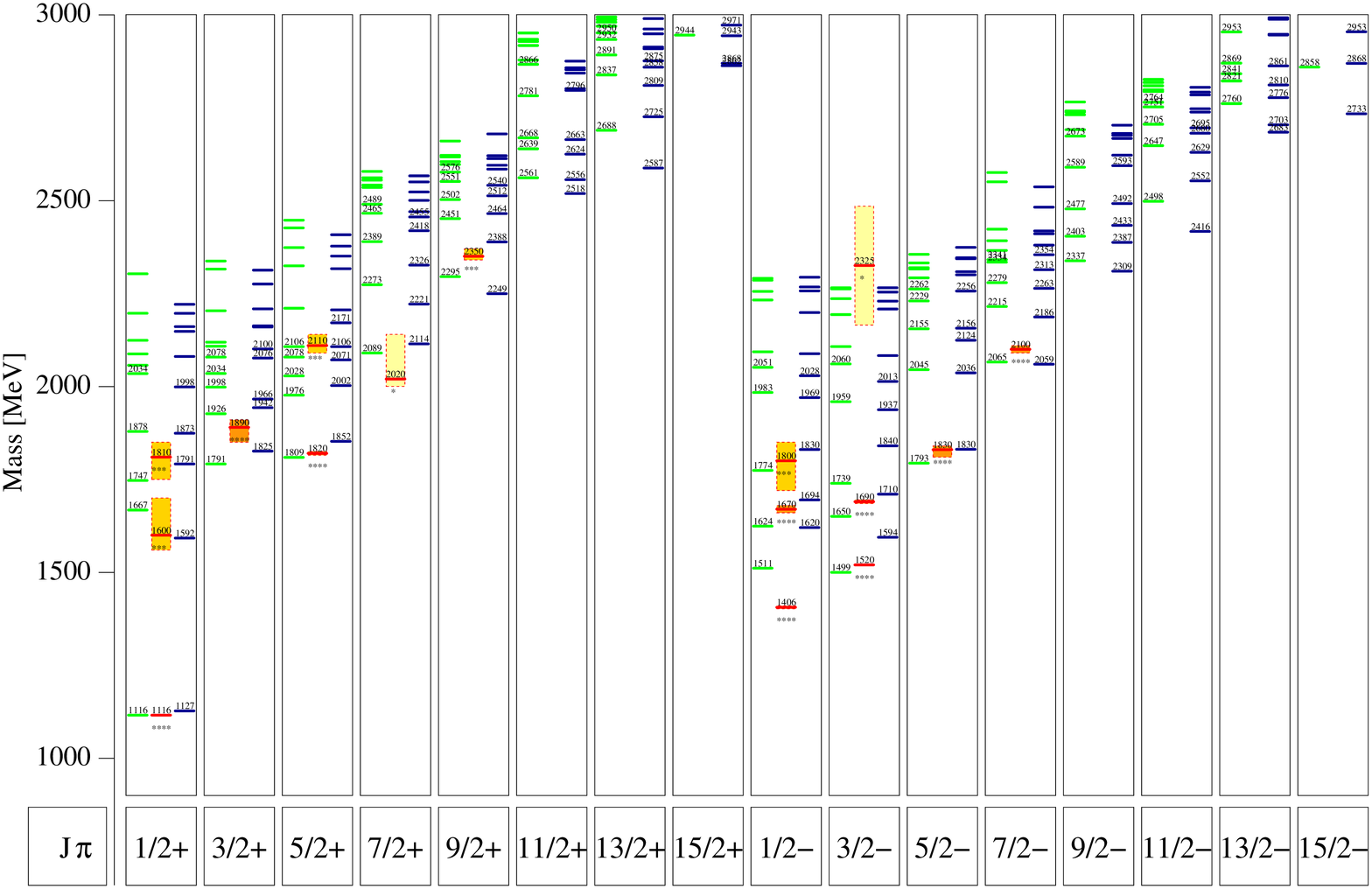}
  \caption{
    Comparison of the $\Lambda$-Spectrum calculated within the present model
    $\mathcal{C}$ (right side of each column) with experimental data from the
    Particle Data Group~\cite{PDG} (central in each column) and with the results
    from model $\mathcal{A}$ of ~\cite{LoeMePe2} (left side in each column), see
    also caption to fig.~\ref{Fig_GaussDeltaPSMeson}.
    \label{Fig_GaussLambdaPSMeson}
  }
\end{figure*}
\begin{figure*}[!htb]
  \centering
  \includegraphics[width=\linewidth]{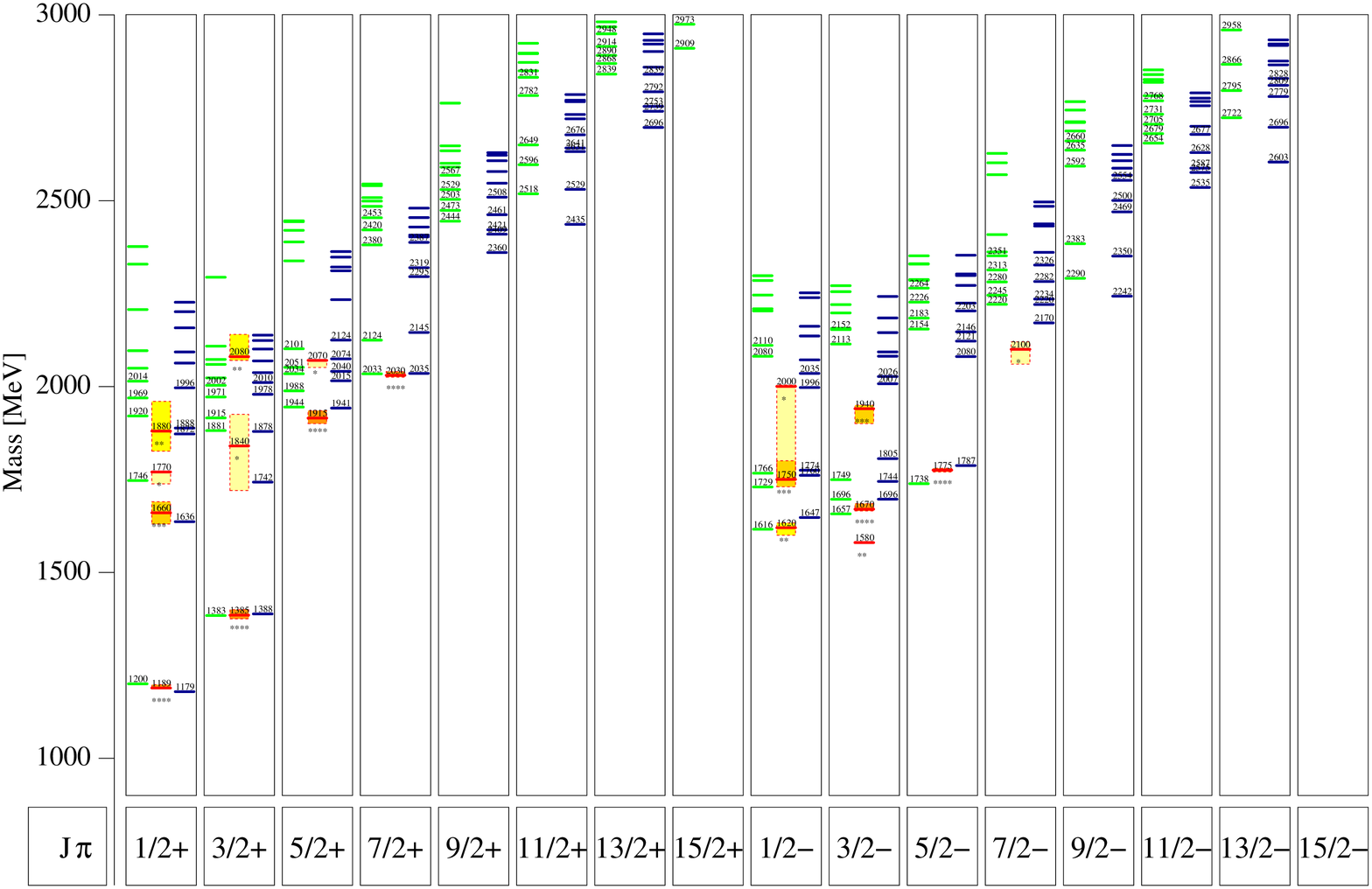}
  \caption{
    Comparison of the $\Sigma$-Spectrum calculated within the present model
    $\mathcal{C}$ (right side of each column) with experimental data from the
    Particle Data Group~\cite{PDG} (central in each column) and with the results
    from model $\mathcal{A}$ of ~\cite{LoeMePe2} (left side in each column), see
    also caption to fig.~\ref{Fig_GaussDeltaPSMeson}.
    \label{Fig_GaussSigmaPSMeson}
  }
\end{figure*}
\begin{figure*}[!htb]
  \centering
  \psfrag{Xi}[c][c]{\Huge$\Xi$}
  \psfrag{Om}[c][c]{\Huge$\Omega$}
  \includegraphics[width=\linewidth]{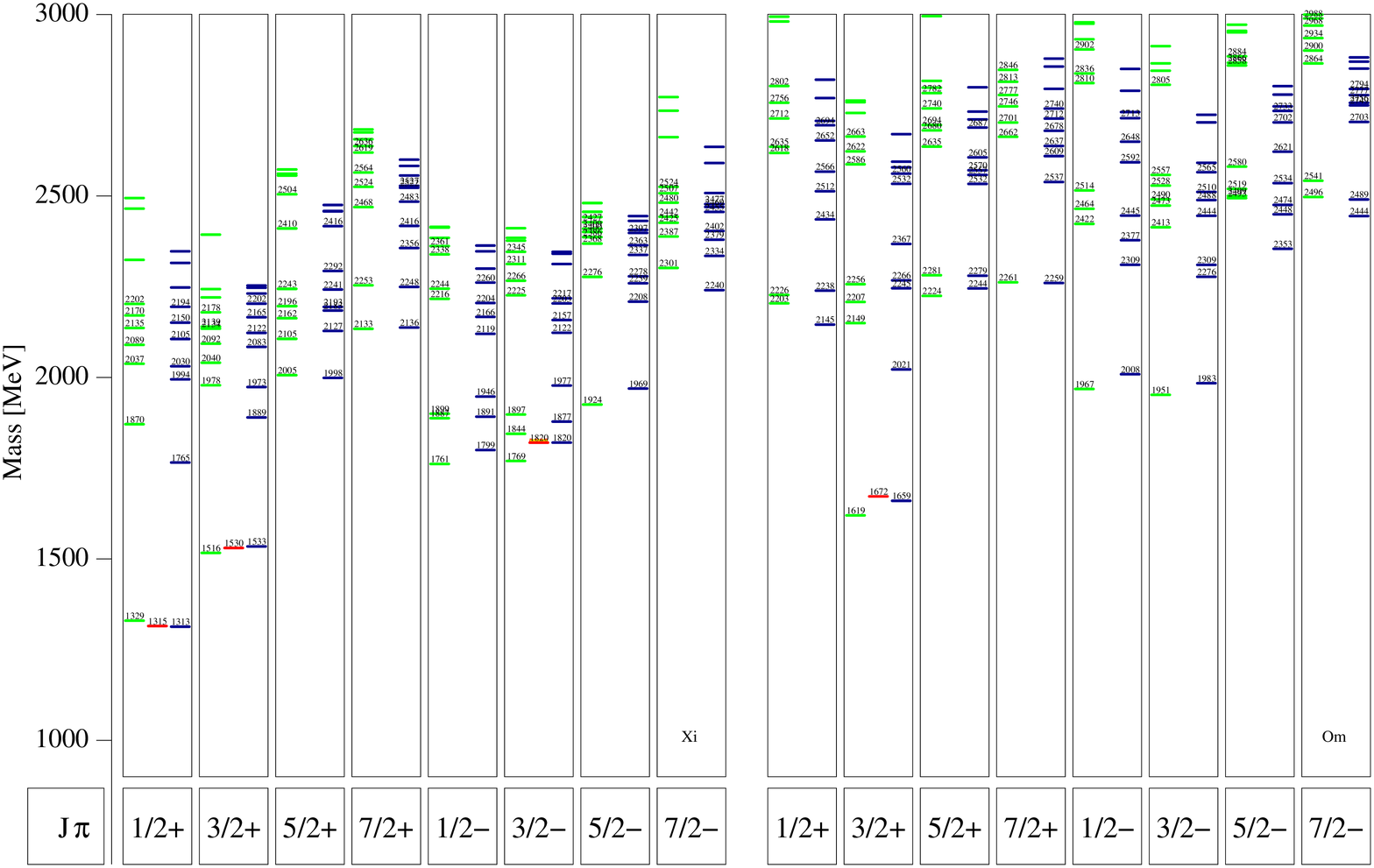}
  \caption{
    Comparison of the $\Xi$-Spectrum (first eight columns) and the
    $\Omega$-Spectrum (rightmost eight columns) calculated within the present model
    $\mathcal{C}$ (right side of each column) with the experimental data from the
    Particle Data Group~\cite{PDG} (central in each column) and with the results
    from model $\mathcal{A}$ of ~\cite{LoeMePe2} (left side in each column), see
    also caption to fig.~\ref{Fig_GaussDeltaPSMeson}.
    \label{Fig_GaussXiOmegaPSMeson}
  }
\end{figure*}

The resulting spectra for hyperon resonances, \emph{viz.} the
$\Lambda$\,,$\Sigma$ and $\Xi$ states are depicted in
figs.~\ref{Fig_GaussLambdaPSMeson}, \ref{Fig_GaussSigmaPSMeson} and
\ref{Fig_GaussXiOmegaPSMeson}\,, respectively.
Again we indeed find an improved description of the ``Roper-like'' resonances
$\Lambda_{\frac{1}{2}^+}(1600)$ and $\Sigma_{\frac{3}{2}^+}(1660)$. Note,
however, that both were used to determine the model parameters. Concerning the
negative parity resonances, although we do find an acceptable description of
the $\Lambda$ resonances with $J^\pi = \frac{3}{2}^-\,, \frac{5}{2}^-$ and
$\frac{7}{2}^-\,$, also the new cal\-culat\-ion can not account for the low
position of the $\Lambda_{\frac{1}{2}^-}(1405)$ resonance, which now is 200
MeV below the calculated position. In our opinion this underlines the
conclusion, that this state cannot indeed be accounted for in terms of a
$q^3$ excitation alone and that its position is determined by a strong
coupling of a ``bare'' $q^3$ state to meson-baryon decay channels due to the
proximity of the $\bar{K}N$-threshold, see also
ref.~\cite{Jido,Hyodo2008_1,Hyodo2008_2} for a description of this state in a
chiral unitary approach.

Concerning the $\Xi$ resonances, with respect to model $\mathcal{A}$
of~\cite{LoeMePe3} mainly the prediction for the excited state
$\Xi(\frac{1}{2}^+)$ at 1765 MeV is 100 MeV lower. To a lesser extend this also
holds for the excited $\Xi(\frac{3}{2}^+)$ which is now predicted at 1889 MeV.

\section{Electromagnetic properties}
\label{sec:electromagneticFF}

As has been elaborated in~\cite{Merten} in lowest order the transition current
matrix element for an initial baryon state with four-momentum 
$\bar{P}'=M=(M,\vec 0)$
in its rest frame and a final baryon state with four-momentum 
$\bar{P}$
is given by the expression
\begin{eqnarray}
  \label{eq:em1}
    &&\langle \bar P | j^\mu(0) | M \rangle
    =
    \!-3\!\!\int\!\!\frac{\textrm{d}^4p_\xi}{(2\pi)^4}\!\!
    \int\!\!\frac{\textrm{d}^4p_\eta}{(2\pi)^4}\!
    \textstyle
    {\bar{\Gamma}}^\Lambda_{\bar{P}}\!\left(p_\xi,p_\eta\!-\!\frac{2}{3}q\right)
  \nonumber
  \\
  &&\hspace*{1em}
  \textstyle
  S^1_F\!\left(\frac{1}{3}M+\!p_\xi\!+\!\frac{1}{2}p_\eta\right)\!
  \otimes\!
  S^2_F\!\left(\frac{1}{3}M-\!p_\xi\!+\!\frac{1}{2}p_\eta\right)
  \nonumber
  \\
  &&\hspace*{1em}
  \textstyle
  \otimes
  S^3_F\!\left(\frac{1}{3}M\!-\!p_\xi\!-\!p_\eta\!+\!q\right)
  \widehat q\gamma^\mu
  S^3_F\!\left(\frac{1}{3}M\!-\!p_\xi\!-\!p_\eta\right)
  \nonumber
  \\
  &&
  \hspace*{10em}
  \textstyle
  \Gamma^\Lambda_M\left(\vec p_\xi,\vec p_\eta\right)\,.
\end{eqnarray}
Here $q$ denotes the momentum transfer, $\widehat q$ is the charge operator
and
\begin{eqnarray}
  \label{eq:em2}
  \lefteqn{
    \Gamma^\Lambda_M\left(\vec p_\xi,\vec p_\eta\right)
    :=
    -\textrm{i}
    \int\frac{\textrm{d}p'_\xi}{(2\pi)^4}
    \int\frac{\textrm{d}p'_\eta}{(2\pi)^4}
  }\nonumber
  \\
  &&
  \hspace*{1em}\left[
    V^{(3)}_\Lambda\left(\vec p_\xi,\vec p_\eta;\vec p'_\xi,\vec p'_\eta\right)
    +
    V^{\textrm{eff}}_\Lambda\left(\vec p_\xi,\vec p_\eta;\vec p'_\xi,\vec
      p'_\eta\right)
  \right]\nonumber
  \\
  &&
  \hspace*{10em}\Phi^\lambda_M\left(\vec p'_\xi,\vec p'_\eta\right)
\end{eqnarray}
is the vertex function in the rest frame of the baryon with $V_\Lambda$ the
projection of the instantaneous interaction kernels onto the subspace of
purely positive and purely negative energy components only, see in particular
Appendix A of~\cite{Merten} for details. The vertex for a general
four-momentum on the mass shell can be obtained by an appropriate
Lorentz-boost.

Accordingly, the Sachs form factors are given by
\begin{subequations}
\begin{eqnarray}
  \label{eq:em3a}
  G^N_E(Q^2) 
  &=& 
  \frac{\langle N, \bar{P}, \frac{1}{2}|j^E_0(0)|N, M,
    \frac{1}{2}\rangle}{\sqrt{4M^2+Q^2}}
  \\
  \label{eq:em3b}
  G^N_M(Q^2) 
  &=& 
  \frac{\langle N, \bar{P}, \frac{1}{2}|j^E_+(0)|N, M,
    -\frac{1}{2}\rangle}{2\sqrt{Q^2}}
\end{eqnarray}
\end{subequations}
where $\left| N, \bar P, \lambda\right\rangle$ denotes a nucleon state
with four-mo\-men\-tum $\bar P$ and helicity $\lambda$. Furthermore
$Q^2 = -q^2$ and $j^E_\pm := j^E_1 \pm \textrm{i}\,j^E_2$\, as well as 
$|\vec P|^2 = Q^2 +$ $\frac{(M^2-{M'}^2-Q^2)^2}{4\,M^2}$\,.
Likewise, the axial vector form factor is given by
\begin{eqnarray}
  \label{eq:em4}
  G_A(Q^2) 
  &=& 
  \frac{\langle p, \bar{P}, \frac{1}{2}|j^A_+(0)|n, M,
    -\frac{1}{2}\rangle}{\sqrt{4M^2+Q^2}}
\end{eqnarray}
where again $j^A_\pm := j^A_1 \pm$ $\textrm{i}\,j^A_2$ with $j^A_\mu$ the
axial current operator, whose matrix elements are given by Eq.~(\ref{eq:em1})
after the formal substitution $\hat q \,\mapsto\,$ $\tau^+\gamma^5$\,.
Of course, the normalisations of the form factors is such that the static
magnetic moments and the axial coupling are given by
\begin{displaymath}
  \mu_M := G_M(Q^2=0)\,,\qquad g_A := G_A(Q^2=0)\,.
\end{displaymath}

\subsection{The electric form factors of the nucleon\label{nucleon_EFF}}
\begin{figure}[!htb]
\centering
\psfrag{x-axis}[c][c]{$Q^2\,\,[\textrm{GeV}^2]$}
\psfrag{y-axis}[c][c]{$G_E^p(Q^2)/G_D(Q^2)$}
\psfrag{MMD}[r][r]{\scriptsize MMD \cite{Mergell}}
\psfrag{Christy}[r][r]{\scriptsize Christy \cite{Christy}}
\psfrag{Qattan}[r][r]{\scriptsize Qattan \cite{Qattan}}
\psfrag{Merten}[r][r]{\scriptsize model $\mathcal{A}$ \cite{Merten}}
\psfrag{Gauss}[r][r]{\scriptsize model $\mathcal{C}$}
\includegraphics[width=\linewidth]{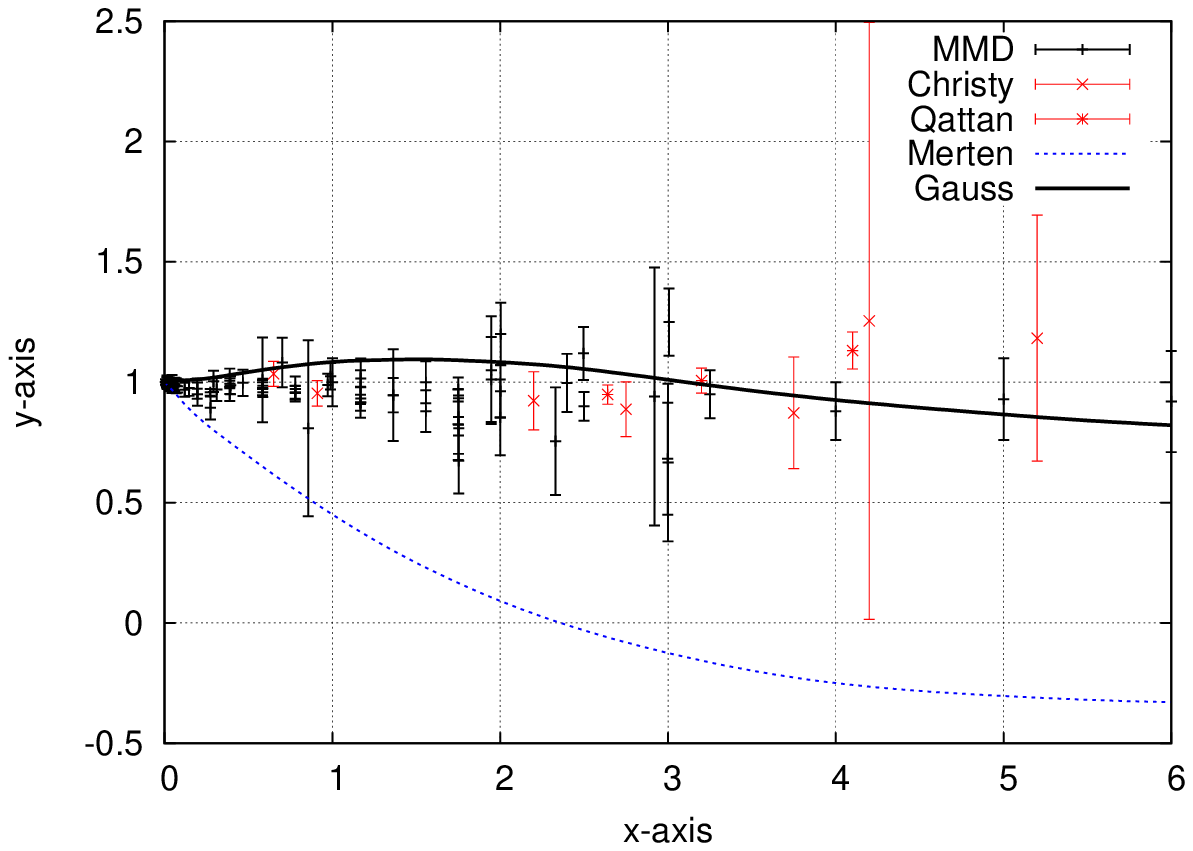}
\caption{
 The electric form factor of the proton divided by the dipole form $G_D(Q^2)$, Eq.~(\ref{eq:EFF}). 
 MMD-Data are taken from Mergell \textit{et al.}~\cite{Mergell}, supplemented by
 data from Christy \textit{et al.}~\cite{Christy} and Qattan \textit{et al.}~\cite{Qattan}\,.
 The solid black line represents the results from the present model
 $\mathcal{C}$;  the dashed blue line those from 
 model $\mathcal{A}$ of~\cite{Merten}, albeit
 recalculated with higher numerical precision. Red data points are taken
 from polarisation experiments and black ones are obtained by Rosenbluth separation.
 \label{EFF_Proton} }
\end{figure}

\begin{figure}[!htb]
\centering
\psfrag{x-axis}[c][c]{$Q^2\,\,[\textrm{GeV}^2]$}
\psfrag{y-axis}[c][c]{$G_E^n(Q^2)$}
\psfrag{MMD}[r][r]{\scriptsize MMD \cite{Mergell}}
\psfrag{Rosenbluth}[r][r]{\scriptsize \cite{Eden,Herberg,Ostrick,Passchier,Schiavilla}}
\psfrag{Polarisation}[r][r]{\scriptsize \cite{Rohe,Golak,Zhu,Madey,Warren,Glazier,Alarcon}}
\psfrag{Merten}[r][r]{\scriptsize model $\mathcal{A}$ \cite{Merten}}
\psfrag{PS-coupl}[r][r]{\scriptsize PS Yukawa}
\psfrag{PV-coupl}[r][r]{\scriptsize PV Yukawa}
\psfrag{Gauss}[r][r]{\scriptsize model $\mathcal{C}$}
\includegraphics[width=1.0\linewidth]{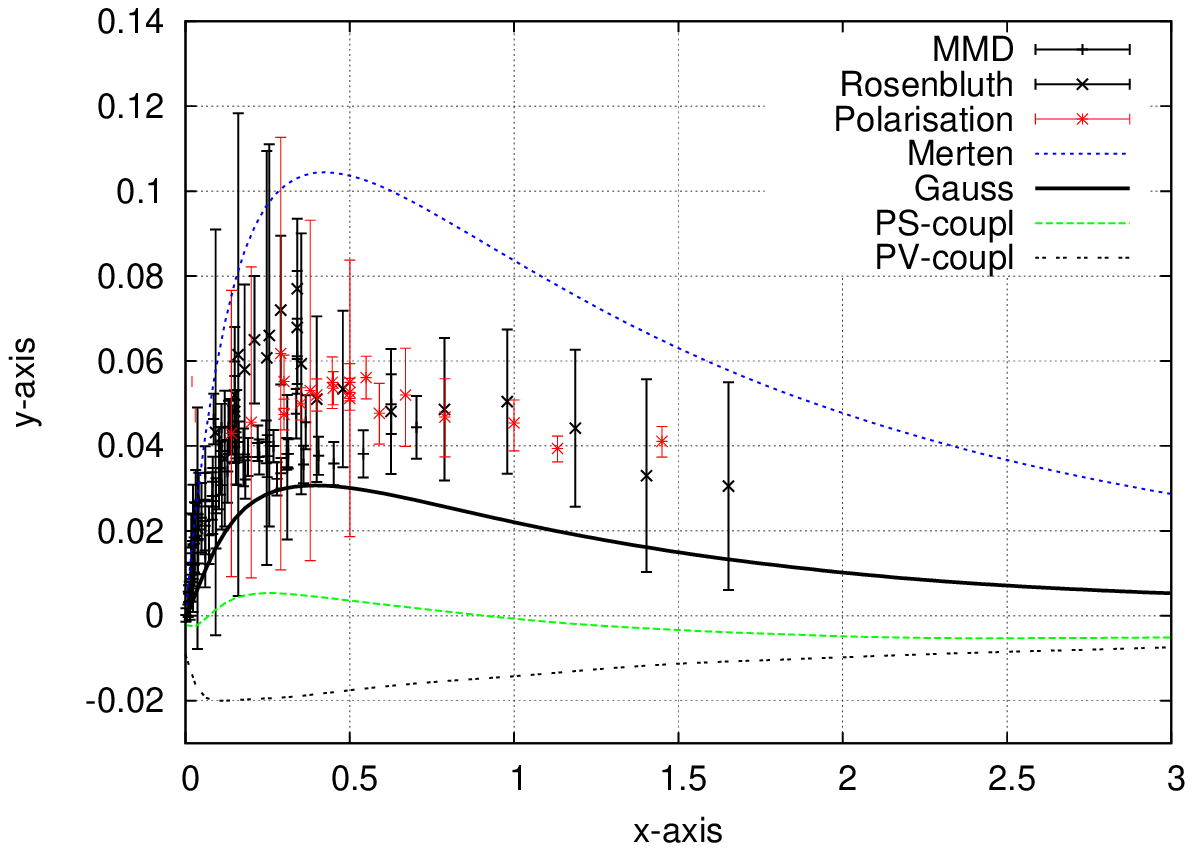}
\caption{The electric form factor of the neutron. MMD-Data are
 taken from the compilation of Mergell \emph{et al.}~\cite{Mergell}.
 The solid black line represents the results from the present model
 $\mathcal{C}$; the green- and brown-dashed lines corresponds to the
 Yukawa-like models with pseudoscalar and/or pseudovector coupling,
 see Eqs.~(\ref{eq:OBEVps}) and~(\ref{eq:OBEVpv}) respectively;
 the dashed blue line in the result from model $\mathcal{A}$
 of~\cite{Merten}, albeit recalculated with higher numerical precision.
 Red data points are taken from polarisation experiments and black
 ones are obtained by Rosenbluth separation.
 \label{EFF_Neutron}}
\end{figure}

In fig.~\ref{EFF_Proton} and \ref{EFF_Neutron} we display the electric proton
and neutron form factor, respectively, up to a momentum transfer of $Q^2=6.0$
$\textrm{GeV}^2$. The black solid curve is the result of the present model
$\mathcal{C}$, the blue dashed curve is the result obtained with the
para\-meters of model $\mathcal{A}$, as in~\cite{Merten}, albeit with a better
numerical precision, see the end of Subsection~\ref{subsec:BSAppr}\,.

Although the electric form factor of the proton, see 
fig.~\ref{EFF_Proton}, 
as calculated with model $\mathcal{A}$ in~\cite{Merten} fell too steeply in
comparison to experimental data, with the present interaction we find a much
improved shape which yields a satisfactory description even up to momentum
transfers of 6~GeV$^2$\,. Indeed, in contrast to model $\mathcal{A}$, which
mainly failed with respect to the isovector part of the form factor, in the
present model $\mathcal{C}$ this form factor shows an almost perfect dipole
shape with the parametrisation
\begin{align}
   \label{eq:EFF}
   G_D(Q^2) = \frac{1}{(1+Q^2/M_V^2)^2}\,,
\end{align}
taken from \cite{Mergell,Bodek} with $M_V^2=0.71\,\textrm{GeV}^2$.

The resulting electric neutron form factor, see
fig.~\ref{EFF_Neutron}, 
has a maximum at approximately the experimental value of $Q^2 \approx 0.4
\textrm{ GeV}^2$ but underestimates the experimental data
from~\cite{Schiavilla} by about the same amount as the earlier calculation
overestimated the data. However, the prediction of model $\mathcal{C}$ is very
similar to the predictions of the Graz group \cite{Plessas} and \cite{Melde07}
for the Goldstone-boson-exchange quark models. The corresponding charge radii
are given in table~\ref{TAB_StatProp}\,. As for the form factor the resulting
squared charge radius of the neutron is calculated too small by a factor of
two. Also the r.m.s. proton radius is slightly smaller than the experimental
value.
\begin{table}[htb]
\centering
\parbox{\linewidth}{
\caption{
  Static properties of the nucleon. The values in parentheses are as reported
  in~\cite{Merten}, the values on top of these are obtained within the same
  model $\mathcal{A}$ but with higher numerical accuracy. The columns PS and
  PV are the results obtained with the additional interaction kernels according
  to Eqs.~(\ref{eq:OBEVps}) and~(\ref{eq:OBEVpv}), respectively, see also
  Section~\ref{sec:Conc} for a brief remark. The static values are extra\-polated
  from a dipole shape-like fit.
  \label{TAB_StatProp}
}
}
\begin{tabular}{lc@{\hspace*{2pt}}c@{\hspace*{2pt}}c@{\hspace*{2pt}}c@{\hspace*{2pt}}c@{\hspace*{2pt}}c@{\hspace*{2pt}}r}
\toprule
                                                                 & Model $\mathcal{A}$   & PS   & PV  & model $\mathcal{C}$  & exp.                   & ref.\\\midrule
\multirow{2}{*}{$\mu_p[\mu_N]\!\!\!\!$}                      & 2.76     & \multirow{2}{*}{2.49} & \multirow{2}{*}{2.39} & \multirow{2}{*}{2.54} &  \multirow{2}{*}{2.793\!\!\!\!}& \multirow{2}{*}{\cite{PDG}}\\
                                                                 & [2.74]      &      &               &                        &                       & \\\midrule
\multirow{2}{*}{$\mu_n[\mu_N]\!\!\!\!$}                      & -1.71                 & \multirow{2}{*}{-1.59}& \multirow{2}{*}{-1.54}& \multirow{2}{*}{-1.59}& \multirow{2}{*}{-1.913\!\!\!\!}& \multirow{2}{*}{\cite{PDG}}\\
                                                                 & [-1.70]             &                        &                        &                       & \\\midrule
\multirow{2}{*}{$\!\!\!\!\sqrt{\langle r^2\rangle_E^p}$[fm]} & 0.91                  & \multirow{2}{*}{0.83} & \multirow{2}{*}{0.68} & \multirow{2}{*}{0.81} & \multirow{2}{*}{0.847\!\!\!\!} & \multirow{2}{*}{\cite{Mergell}}\\
                                                                 & [0.82]             &                        &                        &                       & \\\midrule
\multirow{2}{*}{$\langle r^2\rangle_E^n$[fm]$^2$}            & -0.20                 & \multirow{2}{*}{0.01}& \multirow{2}{*}{0.08}  & \multirow{2}{*}{-0.06}& -0.123\!\!\!\!                  & \multirow{2}{*}{\cite{Mergell}}\\
                                                                 & [-0.11]    &           &                        &                                               & $\pm$0.004             & \\\midrule
\multirow{2}{*}{$\!\!\!\!\sqrt{\langle r^2\rangle_M^p}$[fm]} & 0.90                  & \multirow{2}{*}{0.81} & \multirow{2}{*}{0.68} & \multirow{2}{*}{0.78} & \multirow{2}{*}{0.836\!\!\!\!} & \multirow{2}{*}{\cite{Mergell}}\\
                                                                 & [0.91]    &            &                        &                        &                       & \\\midrule
\multirow{2}{*}{$\!\!\!\!\sqrt{\langle r^2\rangle_M^n}$[fm]} & 0.84                  & \multirow{2}{*}{0.79} & \multirow{2}{*}{0.67}  & \multirow{2}{*}{0.75}  & \multirow{2}{*}{0.889}\!\!\!\! & \multirow{2}{*}{\cite{Mergell}}\\
                                                                 & [0.86]     &           &                        &                        &                       & \\\midrule
\multirow{2}{*}{$g_A$}                                           & 1.22                 & \multirow{2}{*}{1.17}  & \multirow{2}{*}{1.14} & \multirow{2}{*}{1.13}  & 1.267\!\!\!\!                        & \multirow{2}{*}{\cite{PDG,Bodek}}\\
                                                                 & [1.21]    &           &                        &                                               & $\pm$0.0035\!\!\!\!            & \\\midrule
\multirow{2}{*}{$\!\!\sqrt{\langle r^2\rangle_A}$[fm]}       & 0.68                 & \multirow{2}{*}{0.64} & \multirow{2}{*}{0.48}  & \multirow{2}{*}{0.57}  & 0.67\!\!\!\!                   & \multirow{2}{*}{\cite{Bernard}}\\
                                                                 & [0.62]                &                          &                       & &$\pm$0.01\!\!\!\!              & \\\bottomrule
\end{tabular}
\end{table}

\begin{table}[htb]
\centering
\caption{Octet hyperon magnetic moments $\mu$ for model $\mathcal{A}$ and $\mathcal{C}$
calculated as in \cite{Haupt}. The values are given in units of $\mu_N$.\label{TAB_Octet}}
\begin{tabular}{c@{\hspace*{10pt}}c@{\hspace*{10pt}}c@{\hspace*{10pt}}c@{\hspace*{10pt}}}
\toprule
hyperon    & model $\mathcal{A}$ & model $\mathcal{C}$ & PDG \cite{PDG} \\\midrule
$\Lambda$  & -0.606 & -0.577 & -0.613 $\pm$0.004\\\midrule
$\Sigma^+$ &  2.510 &  2.309 &  2.458 $\pm$0.010\\\midrule
$\Sigma^0$ &  0.743 &  0.701 &        -         \\\midrule
$\Sigma^-$ & -1.013 & -0.908 & -1.160 $\pm$0.025\\\midrule
$\Xi^0$    & -1.324 & -1.240 & -1.250 $\pm$0.014\\\midrule
$\Xi^-$    & -0.533 & -0.532 & -0.651$\pm$0.0025\\\bottomrule
\end{tabular}
\end{table}

\begin{table}[htb]
\centering
\caption{Decuplet hyperon magnetic moments $\mu$ for model $\mathcal{A}$ and $\mathcal{C}$
calculated as in \cite{Haupt}. The values are given in units of $\mu_N$.\label{TAB_Decuplet}}
\begin{tabular}{c@{\hspace*{10pt}}c@{\hspace*{10pt}}c@{\hspace*{10pt}}c@{\hspace*{10pt}}}
\toprule
hyperon          & model $\mathcal{A}$ & model $\mathcal{C}$ & PDG \cite{PDG} \\\midrule
$\Delta^{++}$    & 4.241 & 4.238  &  	3.7 to 7.5\\\midrule
$\Delta^+$       & 2.121 & 2.119  & 	$2.7_{-1.3}^{+1.0}\pm1.5\pm3$\\\midrule
$\Delta^0$       & 0.0   & 0.0    &  	\\\midrule
$\Delta^-$       &-2.121 &-2.119  &  	\\\midrule
$\Sigma^{\ast+}$ & 2.567 & 2.431  & 	\\\midrule
$\Sigma^{\ast0}$ & 0.275 & 0.205  & 	\\\midrule
$\Sigma^{\ast-}$ &-2.017 &-2.021  & 	\\\midrule
$\Xi^{\ast0}$    & 0.607 & 0.474  & 	\\\midrule
$\Xi^{\ast-}$    &-1.865 &-1.765  &   	\\\midrule
$\Omega^-$       & -1.675& -1.577 &     $-2.02\pm0.05$\\\bottomrule
\end{tabular}
\end{table}

\begin{figure}[!htb]
\centering
\psfrag{x-axis}[c][c]{$Q^2\,\,[\textrm{GeV}^2]$}
\psfrag{y-axis}[c][c]{$G_M^p(Q^2)/G_D(Q^2)/\mu_p$}
\psfrag{MMD}[r][r]{\scriptsize MMD \cite{Mergell}}
\psfrag{Christy}[r][r]{\scriptsize Christy \cite{Christy}}
\psfrag{Qattan}[r][r]{\scriptsize Qattan \cite{Qattan}}
\psfrag{Bartel}[r][r]{\scriptsize Bartel \cite{Bartel}}
\psfrag{Merten}[r][r]{\scriptsize Merten $\mathcal{A}$ \cite{Merten}}
\psfrag{Gauss}[r][r]{\scriptsize model $\mathcal{C}$}
\includegraphics[width=1.0\linewidth]{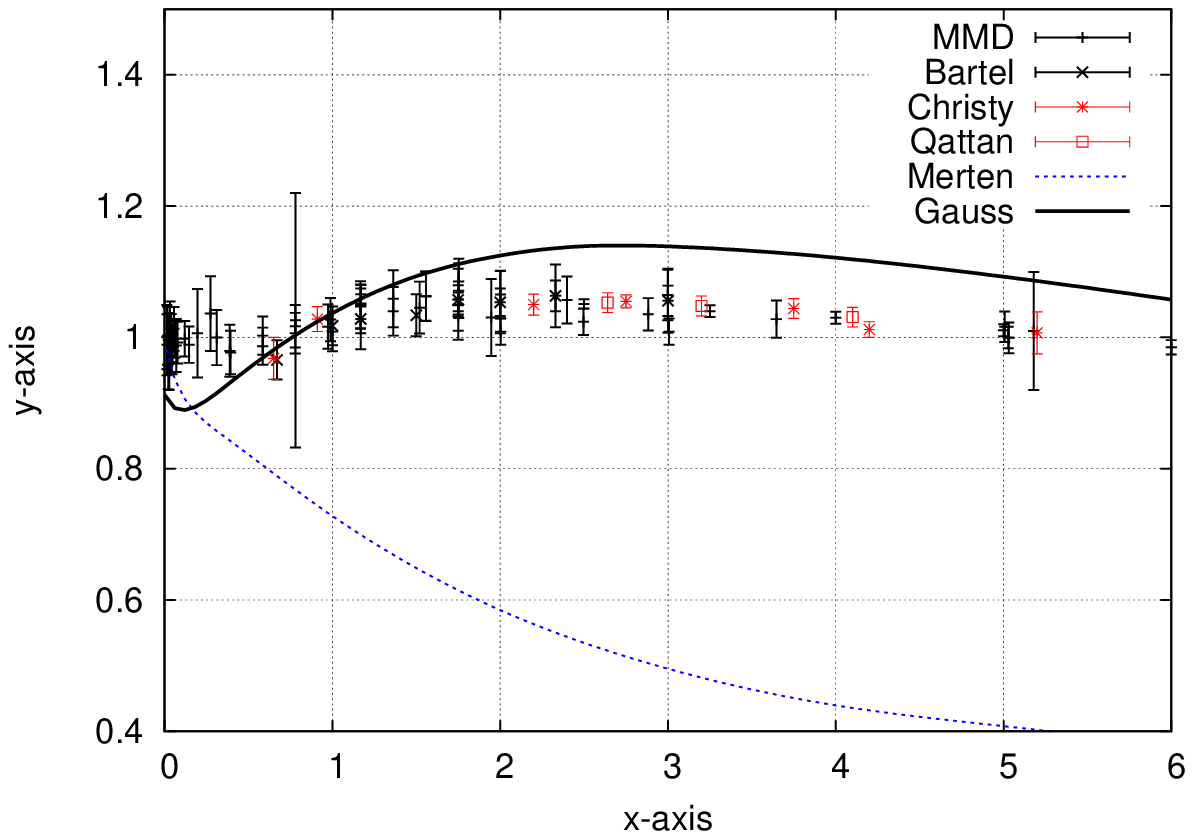}
\caption{The magnetic form factor of the proton divided by the dipole form
 $G_D(Q^2)$, Eq.~(\ref{eq:EFF}) and the magnetic moment of the
 proton $\mu_p=2.793\,\mu_N$. MMD-Data are taken from the compilation
 of Mergell \emph{et al.}~\cite{Mergell}. Additionally, polarisation
 experiments are marked by red. The black marked data points are obtained by
 Rosenbluth separation.\label{MFF_Proton}}
\end{figure}

\begin{figure}[!htb]
\centering
\psfrag{x-axis}[c][c]{$Q^2\,\,[\textrm{GeV}^2]$}
\psfrag{y-axis}[c][c]{$G_M^n(Q^2)/G_D(Q^2)/\mu_n$}
\psfrag{MMD}[r][r]{\scriptsize MMD \cite{Mergell}}
\psfrag{Anklin}[r][r]{\scriptsize Anklin \cite{Anklin}}
\psfrag{Kubon}[r][r]{\scriptsize Kubon \cite{Kubon}}
\psfrag{Xu}[r][r]{\scriptsize Xu \cite{Xu}}
\psfrag{Madey}[r][r]{\scriptsize Madey \cite{Madey}}
\psfrag{Alarcon}[r][r]{\scriptsize Alarcon \cite{Alarcon}}
\psfrag{Merten}[r][r]{\scriptsize model $\mathcal{A}$ \cite{Merten}}
\psfrag{Gauss}[r][r]{\scriptsize model $\mathcal{C}$}
\includegraphics[width=1.0\linewidth]{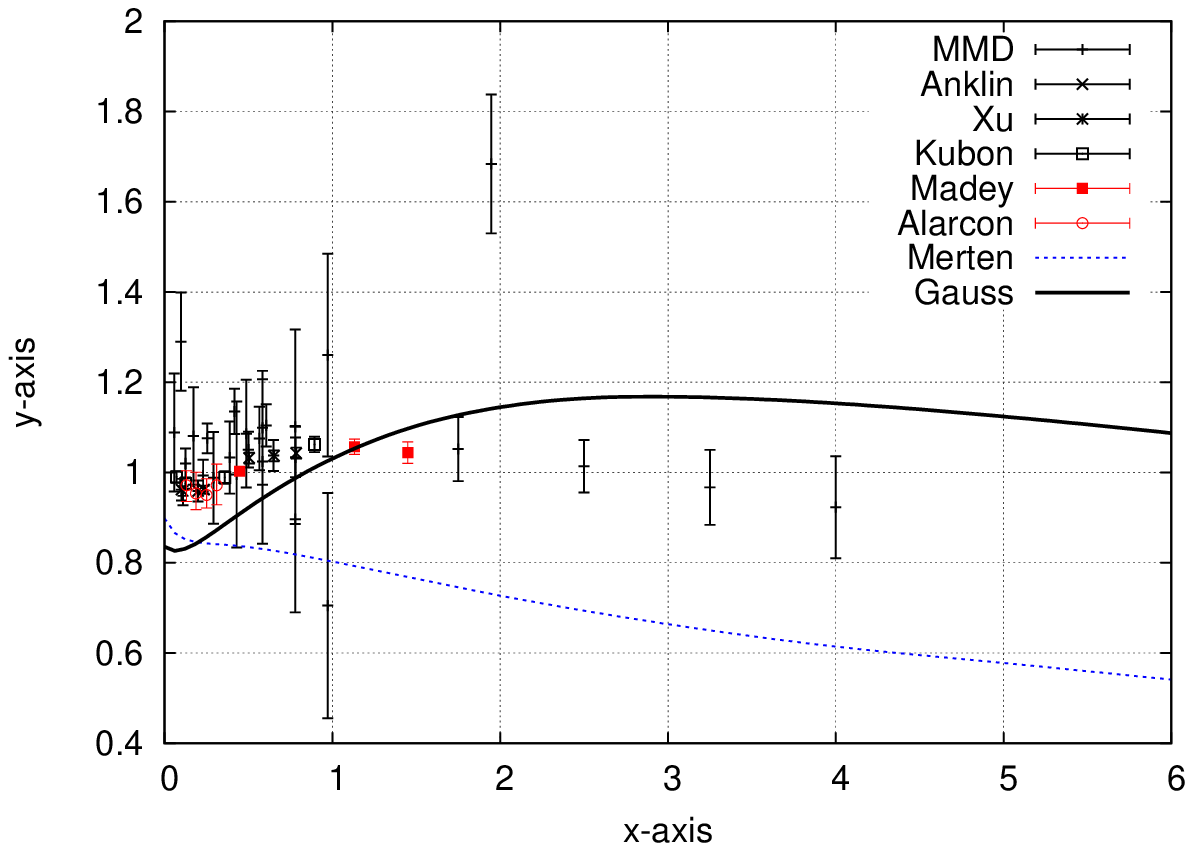}
\caption{The magnetic form factor of the neutron divided by the dipole form $G_D(Q^2)$,
 Eq.~(\ref{eq:EFF}) and the magnetic moment of the neutron $\mu_n=-1.913\,\mu_N$.
 MMD-Data are taken from the compilation by Mergell \emph{et al.}~\cite{Mergell} and
 from more recent results from MAMI~\cite{Anklin,Kubon}\,. Additionally, polarisation
 experiments are marked by red data points. The black marked ones are obtained by
 Rosenbluth separation.\label{MFF_Neutron}}
\end{figure}

In fig.~\ref{MFF_Proton} and ~\ref{MFF_Neutron} we display the magnetic
proton- and neutron form factor up to a momentum transfer of $Q^2=6.0\, 
\textrm{GeV}^2$\,, respectively. 
Again, the black solid curve is the result of the present model $\mathcal{C}$, the blue dashed
curve is the result obtained with the parameters of model $\mathcal{A}$, as
in~\cite{Merten}, albeit with a better numerical precision, see the end of
Subsection~\ref{subsec:BSAppr}\,.

\begin{figure}[!htb]
\centering
\psfrag{x-axis}[c][c]{$Q^2\,\,[\textrm{GeV}^2]$}
\psfrag{y-axis}[c][c]{$\mu_p\,G_E^p / G_M^p(Q^2)$}
\psfrag{Rosenbluth}[r][r]{\scriptsize \cite{Price,Bartel,Berger_1,Walker,Andivahis}}
\psfrag{Polarisation}[r][r]{\scriptsize \cite{Milbrath,Jones,Gayou2001,Pospischil,Gayou2002,Punjabi,Higinbotham,Hu,Crawford,Ron,Zhan}}
\psfrag{Merten}[r][r]{\scriptsize model $\mathcal{A}$ \cite{Merten}}
\psfrag{Gauss}[r][r]{\scriptsize model $\mathcal{C}$}
\includegraphics[width=1.0\linewidth]{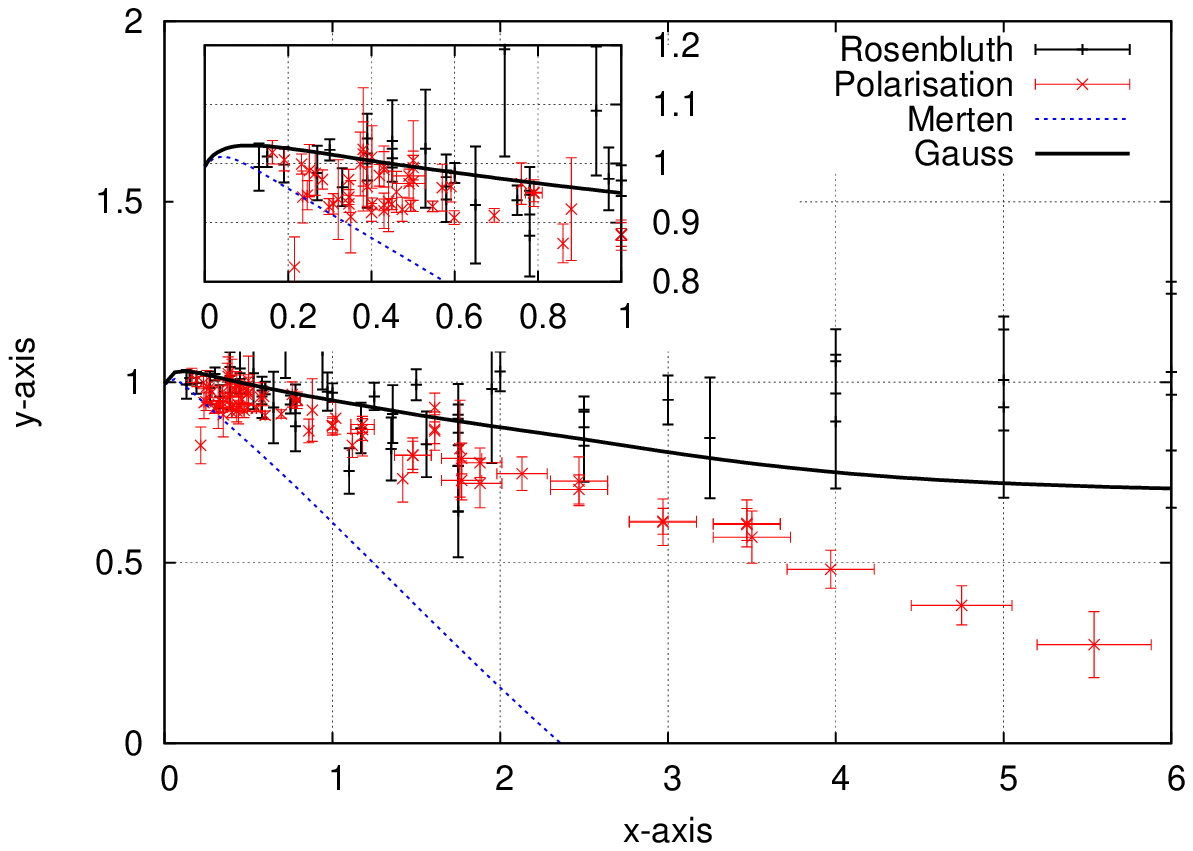}
\caption{The ratio $\mu_p\,G_E^p/G_M^p$ compared to recent JLAB data (see
 legend).
 In the insert the low momentum transfer region is enlarged. The solid
 black line is the present result, the dashed blue line the
 result in model $\mathcal{A}$\,. Red data points are taken from polarisation
 experiments and the black ones are obtained from Rosenbluth separation.
 \label{GEpGMp}}
\end{figure}

Whereas in the original calculation (model $\mathcal{A}$ of~\cite{Merten}) the
absolute value of these form factors dropped slightly too fast as a function
of the momentum transfer, in the present calculation we now find a very good
description even at the highest momentum transfers. Only at low momentum
transfer the values are too small as is reflected by the rather small values
for the various magnetic radii, see table~\ref{TAB_StatProp} and the too small
values of the calculated magnetic moments. Note, however that the ratio
$\mu_p/\mu_n \approx 1.597$ for model $\mathcal{C}$ slightly changes (previously
$\mu_p/\mu_n \approx 1.605$ for model $\mathcal{A}$) and is slightly larger
than the experimental value $\mu_p/\mu_n \approx 1.46$\,; all values are
remarkably close to the non-relativistic constituent quark model value
$\mu_p/\mu_n = \frac{3}{2}$\,. The magnetic moments of flavour octet and decuplet
baryons has been calculated accordingly to the method outlined in \cite{Haupt}. The
results are compared to experimental values in Table \ref{TAB_Octet} and
\ref{TAB_Decuplet}, respectively. As a consequence of the better description
of the momentum transfer dependencies in the individual form factors, we now also
find an improved description of the momentum transfer dependence of the form factor
ratio $\mu_p\,G^p_E / G^p_M(Q^2)$\,, which has been the focus on the discussion
whether two-photon amplitudes are relevant for the discrepancy~\cite{Vanderhaeghen}
found between recent measurements based on polarisation data (red data points of
fig.~\ref{GEpGMp})~\cite{Milbrath,Jones,Gayou2001,Gayou2002,Punjabi,Hu,Crawford,Higinbotham,Ron,Zhan}
versus the traditional Rosenbluth separation (black data points of fig.~\ref{GEpGMp}),
see \emph{e.g.}~\cite{Price,Bartel,Berger_1,Walker}. Whereas in 
the original model $\mathcal{A}$ this ratio fell much too steep, we now find
in model $\mathcal{C}$ a much better description of this quantity, see
fig.~\ref{GEpGMp} for a comparison with various data. Up to $Q^2 \approx 3\,\textrm{GeV}^2$
we indeed find the observed linear dependence.

\begin{figure}[!htb]
\centering
\psfrag{x-axis}[c][c]{$Q^2\,\,[\textrm{GeV}^2]$}
\psfrag{y-axis}[c][c]{$G_A^3(Q^2)/G_D^A(Q^2)/g_A$}
\psfrag{Amaldi}[r][r]{\scriptsize Amaldi \cite{Amaldi}}
\psfrag{Brauel}[r][r]{\scriptsize Brauel \cite{Brauel}}
\psfrag{Bloom}[r][r]{\scriptsize Bloom \cite{Bloom}}
\psfrag{Del Guerra}[r][r]{\scriptsize Del Guerra \cite{DGuerra}}
\psfrag{Joos}[r][r]{\scriptsize Joos \cite{Joos}}
\psfrag{Baker}[r][r]{\scriptsize Baker \cite{Baker}}
\psfrag{Miller}[r][r]{\scriptsize Miller \cite{Miller}}
\psfrag{Kitagaki83}[r][r]{\scriptsize Kitagaki 83 \cite{Kitagaki83}}
\psfrag{Kitagaki90}[r][r]{\scriptsize Kitagaki 90 \cite{Kitagaki90}}
\psfrag{Allasia}[r][r]{\scriptsize Allasia \cite{Allasia}}
\psfrag{Merten}[r][r]{\scriptsize model $\mathcal{A}$ \cite{Merten}}
\psfrag{Gauss}[r][r]{\scriptsize model $\mathcal{C}$}
\includegraphics[width=1.0\linewidth]{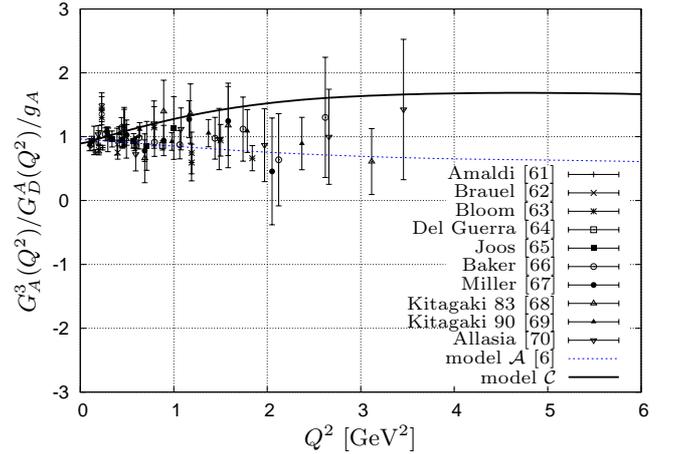}
\caption{The axial form factor of the nucleon divided by the axial dipole
 form in Eq.~(\ref{eq:AFF}) and the axial coupling $g_A=-1.267$. The black solid
 line is the present result, the blue dashed line the
 result in model $\mathcal{A}$\,. Experimental
 data are taken from the compilation by Bernard \emph{et al.}~\cite{Bernard}\,
 \label{AFF_3}\,.} 
\end{figure}

Finally, the axial form factor, see fig.~\ref{AFF_3}, was already very well
described in model $\mathcal{A}$ of~\cite{Merten}\,. Although falling slightly
less steeply, the present calculation still gives a very satisfactory
description of the data also at higher momentum transfers in the same manner
as in~\cite{Glozman2001,Wagenbrunn2001,Wagenbrunn2003}. As for the magnetic
moments the value of the axial coupling constant is too small, but of course
much better than the non-relativistic constituent quark model result $g_A=\frac{5}{3}$\,.
The axial form factor, presented in fig.~\ref{AFF_3}, is divided by the axial dipole form
\begin{align}
   \label{eq:AFF}
   G^A_D(Q^2)=\frac{g_A}{(1+Q^2/M_A^2)^2}\,,
\end{align}
with the parameters $M_A=1.014\pm0.014\,\textrm{GeV}$
and $g_A=1.267$ taken from Bodek \textit{et al.} \cite{Bodek}.

In summary we find, that the new model $\mathcal{C}$, apart from some
improvements in the description of the excitation\,\,\,\, spec\-tra at the expense of
additional parameters of a phenomenologically introduced flavour dependent
interaction does allow for a parameter-free description of electromagnetic
gr\-ound state properties of a similar overall quality as has been obtained
before, with some distinctive improvements on the momentum transfer dependence
of various form factors. A discussion of the momentum dependence of helicity
amplitudes for the electromagnetic excitation of baryon resonances will be
given in a subsequent paper~\cite{Ronniger2}\,.

\section{Summary and conclusion}
\label{sec:Conc}

In the present paper we have tried to demonstrate that by introducing an
additional flavour dependent interaction, para\-metrised with a Gaussian radial
dependence with an universal range and two couplings for flavour octet and
flavour singlet exchange, it is possible to improve upon some deficiencies
found in a former relativistically covariant constituent quark model treatment
of baryonic excitation spectra based on (an instantaneous formulation of) the
Bethe-Salpeter equation. These improvements include:
\begin{itemize}
\item 
  A better description of excited negative parity states sli\-ght\-ly below $2$
  GeV in the $\Delta$ spectrum;
\item
  A better description of the position of the first scalar, iso\-scalar
  excitation of the ground state in all light-fla\-vour sectors;
\item
  An improved description of the momentum de\-pend\-ence of electromagnetic form
  factors of ground states without the introduction of any additional
  parameters.
\end{itemize}
It must be conceded that this additional interaction was introduced purely
phenomenologically and required a\linebreak drastic modification of the parametrisation
of confinement and the other flavour dependent interaction of the original model, which
had a form as inferred from instanton effects. In spite of this, with only 10
parameters in total we still consider this to be an effective description of
the multitude of resonances found for baryons made out of light flavoured
quarks. 
\begin{figure*}[!htb]
  \centering
  \psfrag{De}[c][c]{\Huge$\Delta$}
  \psfrag{Nu}[c][c]{\Huge$N$}
  \includegraphics[width=0.8\linewidth]{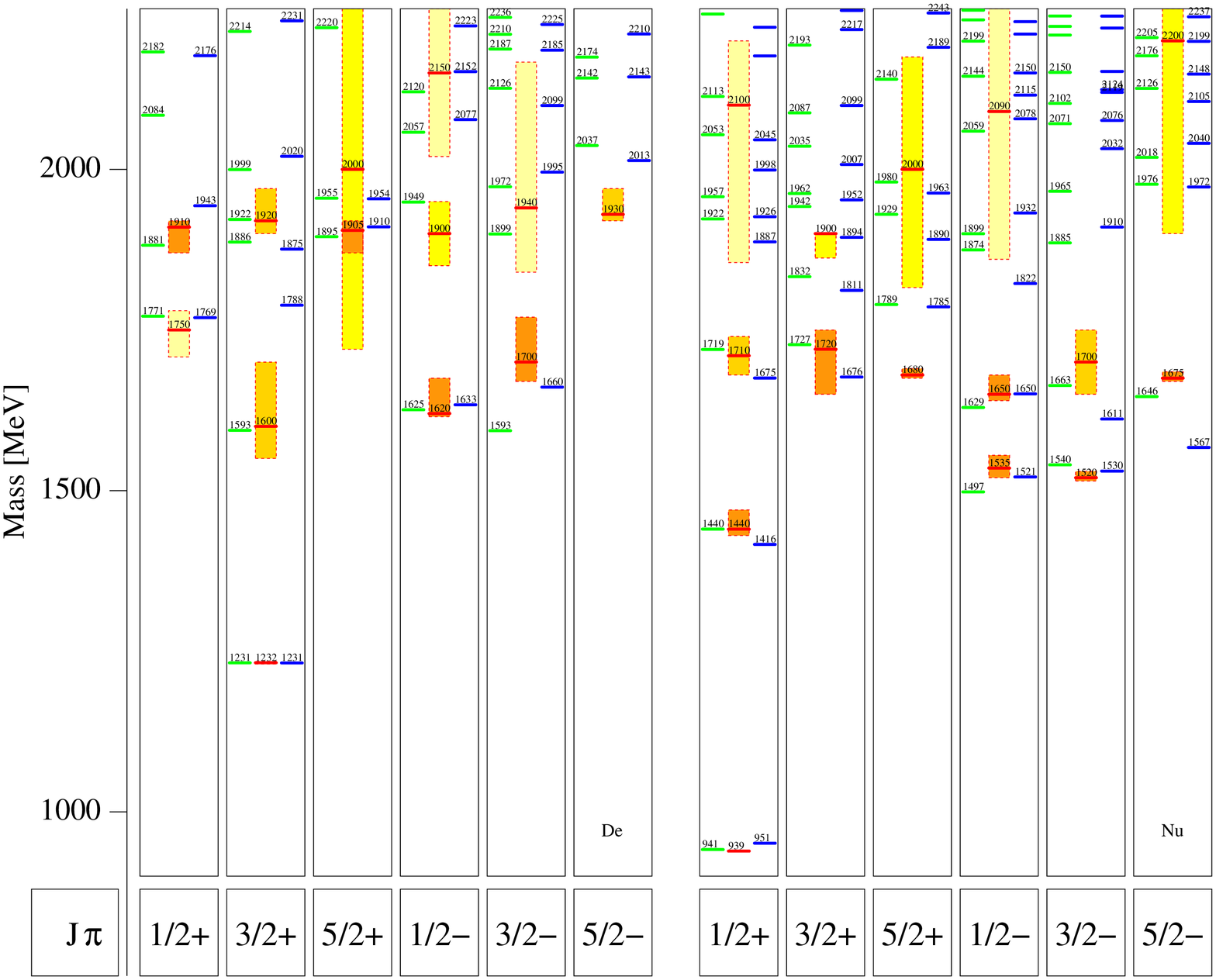}
  \parbox{\linewidth}{
  \caption{
    Comparison of the $\Delta$-Spectrum (first six columns) and the
    $N$-Spectrum (rightmost six columns) calculated within the PS and PV coupled models
    from Eqs.~(\ref{eq:OBEVps}) and~(\ref{eq:OBEVpv}) (right and left side of
    each column) and with the experimental data from the Particle Data Group~\cite{PDG}
    (central in each column).
    \label{Summary_Spec}
  }
  }
\end{figure*}

Nevertheless, it would have been preferred, if the additional flavour dependent
interaction could be related to a genuine physical process, such as light
pseudoscalar meson exchange. Indeed, we tried to find parametrisations of
confinement and parameters of the instanton induced interaction, that could be
combined with interaction kernels as given by the expressions in
Eq.~(\ref{eq:OBEVps}) or~(\ref{eq:OBEVpv}). However, for pseudoscalar
coupling (PS) of a meson nonet to the quarks, we could find a description of the
mass spectra with similar features and of a similar quality as in the present
model $\mathcal{C}$ only at the expense of introducing flavour SU(3) symmetry
breaking and thus introducing more parameters. The latter was also found to be
the case for pseudovector coupling (PV) where, moreover, no significant improvement
concerning the deficiencies in the spectra mentioned above was
found. For the sake of completeness, the nucleon spectrum for PS and PV coupled models
from Eqs.~(\ref{eq:OBEVps}) and~(\ref{eq:OBEVpv}) is shown in fig.~\ref{Summary_Spec}.
Furthermore, with both \emph{An\-satze} we were not able to reproduce in
particular the electric neutron form factor, see fig.~\ref{EFF_Neutron} and
table~\ref{TAB_StatProp} for some typical results. Accordingly, we dismissed
these possibilities and preferred the phenomenological approach of model
$\mathcal{C}$ dis\-cus\-sed in the paper.

A parameter-free calculation of longitudinal and transverse helicity amplitudes
for electro-excitation is presently performed and the results will be dis\-cussed
in a subsequent paper~\cite{Ronniger2}\,.

\appendix

\section*{Acknowledgments}
Stimulating discussions with E. Klempt, A. V. Sarantsev and U. Thoma within the framework of
the DFG supported Collaborative Research Centre SFB/TR16 are gratefully acknowledged.

\end{document}